\documentclass[reprint,amsmath,amssymb,aps,prapplied,floatfix]{revtex4-2}
\makeatletter
\usepackage[hidelinks,colorlinks=true,linkcolor=blue,citecolor=blue, urlcolor = blue]{hyperref}
\usepackage{bookmark}
\newcommand*{\rom}[1]{\expandafter\@slowromancap\romannumeral #1@}
\usepackage{parskip}
\usepackage{graphicx}
\usepackage{dcolumn}
\usepackage{bm}
\usepackage{comment}
\usepackage[normalem]{ulem}
\usepackage[utf8]{inputenc}
\usepackage{hyperref}
\usepackage{lineno}

\usepackage{cleveref}
\usepackage{feynmp-auto}
\usepackage{pgfplots}
\pgfplotsset{compat=1.15}
\usepackage{mathrsfs}
\usetikzlibrary{arrows}
\begin{document}
\preprint{APS/123-QED}
\title{ Jet substructure in neutral current deep inelastic \texorpdfstring{$e^{+}p$}{TEXT} scattering at upcoming Electron-Ion Collider}
\author {Siddharth Jain}
\email{siddharth2006163@st.jmi.ac.in}
\affiliation{Department of Physics, Jamia Millia Islamia University, Delhi, 110025, India}
\author {R. Aggarwal}
\email{ritu.aggarwal1@gmail.com}
\affiliation{USAR, Guru Gobind Singh Indraprastha University, East Delhi Campus, 110092, India}
\author {M. Kaur}
\email{manjit@pu.ac.in}
\affiliation{Department of Physics, Panjab University, Chandigarh 160014, India\\
Department of Physics, Amity University, Punjab, Mohali {140306}, India}
\date{\today}
\begin{abstract}
Predictions are made for the jet substructure of one-jet events produced in positron-proton neutral current deep inelastic scattering at the future Electron-Ion Collider for exchanged four-momentum squared, $Q^2 > 125$ GeV$^2$.~Data are simulated using Monte Carlo event generators PYTHIA 8.304 and RAPGAP 3.308 at the center of mass energies $\sqrt{s}$ = 63.2, 104.9 and 141 GeV.~Jets and subjets are produced by longitudinally invariant $k_T$ cluster algorithm.~The subjet multiplicity distributions and differential jet shapes are measured for different jet sizes and varying jet-resolution parameter.~A comparison is presented between the $k_T$ and anti-$k_T$ cluster algorithms for the study of jets and subjets using the simulated data and the HERA data.
\end{abstract}
\maketitle
\section{\label{sec:level1}INTRODUCTION}
Partons produced in high energy collisions fragment into hadrons.~The process called hadronization corresponds to the transition from partons-to-hadrons.~The hadrons are observed in particle detectors as collimated sprays of particles called jets \cite{klaus2018jet}.~These jets retain information on the underlying partonic interactions which can be used to study partons undergoing hadronization.~The properties of jets are affected by perturbative radiation as well as by low-$p_{T}$ non-perturbative effects.~The description of jet production in high energy collisions is described by quantum chromodynamics (QCD).~As the quantum field theory is not yet fully defined, QCD continues to be guided by experimental observations.~Whether or not, the particle jets of perturbative QCD would exhibit the known properties at high energies at the future colliders with enhanced luminosity, needs investigation.~The evolution of parton radiation within a jet is dictated by QCD by the splitting functions which are calculable as power series in the coupling constant $\alpha_{s}$.~These can be used to exploit the perturbative QCD description of the jet structure in deep inelastic scattering~(DIS) by allowing the measurement of $\alpha_{s}$.~Thus the study of jet production in neutral current~(NC) DIS \cite{roberts1993structure} events provides a rich testing ground for pQCD and parameterisation of the proton parton distribution functions~(PDFs) \cite{klasen2017nuclear}.

In addition the analysis of jet substructure \cite{marzani2019looking,kogler2021advances} also provides information on the transition between a parton produced in a hard process and the experimentally observable jet of hadrons \cite{ellis1992jets}.~The jet-like substructures within a jet are known as subjets and their multiplicity is measured as the number of clusters resolved in a jet.~Characteristics and topology of substructure provide direct access to the QCD splitting functions.~For an inclusive sample of jets produced in NC DIS with high jet transverse energy, the fragmentation effects become negligible, making the subjet multiplicity and its mean, calculable in pQCD.~Further, the dependence of the pQCD calculations on the PDFs is reduced for this variable.~The lowest order contribution to this variable is of the order of first power in $\alpha_{s}$, thus making the stringent test of pQCD beyond leading order possible.~In DIS, the jet production is governed by leading order QCD.~But to describe internal structure of jets, information at the next-to-leading-order~(NLO) is required.

Internal structure of jets has been studied in $e^{+}e^{-}$ collisions at the Large Electron–Positron Collider~(LEP) \cite{lep,aleph2000measurements,delphi1998investigation}, $p\bar{p}$ collisions at Tevatron \cite{tev1,tev2,tev3}, $pp$ collisions at LHC \cite{kaur2013subjet,aad2011study,altheimer2012jet} and $ep$ scattering at HERA \cite{zeus1999measurement}.
Subjet multiplicity and jet shape variables were used to study jet substructure in DIS.~The subjet multiplicity has been used to calculate the strong coupling constant $\alpha_{s}$ \cite{zeusmain, as1, as2}, one of the fundamental parameters of the standard model. 

 Previous studies revealed that jets in DIS events at HERA were similar to those in $e^{+}e^{-}$ events and narrower than in $p\bar{p}$ interactions \cite{opal1998multiplicity,akers1994qcd}.~In NC DIS, the jet became narrower \cite{zeus1999measurement} and resolved to fewer subjets \cite{zeusmain} as the jet transverse energy $E_{T,jet}$ increased.~In photo-production, the mean subjet multiplicity increased and jet shape broadened as the jet pseudorapidity $\eta_{T,jet}$ increased \cite{glasman2001jet}.

Electron-Ion Collider (EIC) is scheduled to be built at Brookhaven National Laboratory~(BNL) in the US. It aims to accelerate polarized nucleon beams which are crucial to study the properties of quarks and gluons in nuclear medium through deep inelastic scattering process. The lepton probe would provide high precision data to explore the hadron structure. The EIC aims to provide a high luminosity of order 10$^{32}$-10$^{34}$ cm$^{-2}$ s$^{-1}$ for $ep$ collisions and 10$^{30}$-10$^{32}$ cm$^{-2}$ s$^{-1}$ for $eAu$ collisions \cite{EIC1}. A minimum center of mass energy (cms) of about 10 GeV is needed for close interaction with quarks and cms energy of order 100 GeV is required to avail higher $Q^{2}$ \cite{EIC2}. At EIC, beam energies would range from 5 to 20 GeV for $e^{+}$ (or $e^{-}$) beam and 25 to 275 GeV for beams of nuclei such as gold, lead or uranium.~The corresponding center of mass energies would vary from 20 to 141 GeV \cite{khalek2022snowmass}. The energy variations would increase the sensitivity to gluon distributions and spin-polarized beams would also facilitate the study of the spin structure of proton \cite{eic_whitepaper}.~Utilizing the proposed variability in the center-of-mass energy in the EIC and particle identification capabilities of its detectors, one can study hadronization within jets in a wide kinematic regime by characterising the $x$ and $Q^2$ scales of the process.

Recently jet studies at the EIC energies have gained momentum and we present here a brief review of different interesting studies performed.~Jets have been studied in photo-production events \cite{aschenauer2020jet} for the $e^+p$ and $e^+A$ simulated data from PYTHIA 6 \cite{pythia6main}. Here it was concluded that in spite of low transverse momentum and small particle multiplicities, the jet substructure study would be feasible at the EIC in the photo-production. Photon-proton processes are generally categorized by the virtuality $Q^2$ of the photon. The value of $Q^2$ is close to zero for quasi-real photo-production interactions whereas, scattering which involves a $Q^2$ larger than a few GeV$^2$ is called DIS. The study by B. Page et al~\cite{page2020experimental} emphasized the importance of jets as probes and gives an example of studying the gluon helicity contributing to the spin of proton through di-jet events in DIS at the future EIC.~In another recent publication by Z.-B. Kang et al~\cite{kang2020jet}, the jet charge has been studied to tag the jet in terms of initiating parton flavors. This information can be used to explore the nucleon flavor and spin structure. The study \cite{arratia2021charm} has explored the feasibility of charm-jet cross sections for charged current DIS interactions at the future EIC. The charm-jet cross sections are known to be sensitive to the strange sea content of the proton. Jets would be used as the precision probes at the future EIC which will be collecting the first ever data from lepton-nucleus collisions.~The data is expected to be cleaner than the ones collected using hadron beam probes. The properties of DIS jets, such as momentum balance and azimuthal correlation with electrons and substructure in terms of number of constituents are presented in detail in \cite{arratia2020jets}. This paper emphasized the need to revisit the HERA data, the only lepton-hadron collider, in the light of future EIC.

The perturbative QCD calculations done by using jet shape and the jet substructure observables known as jet angularities have been compared with the data from HERA, for the process of photoproduction\cite{aschenauer2020jet}. Overall, a good agreement  is shown and it is concluded that jet substructure studies are feasible at EIC despite the relatively low jet transverse momentum and particle multiplicities. Precision jet substructure measurements at the EIC can be used for the tuning of Monte Carlo event generators and the extraction of nonperturbative corrections.

Recently some other techniques have also been developed to improve the understanding of jet substructure. These include the techniques such as jet grooming which is designed to remove soft wide-angle radiation from the identified jets. It allows for a more direct comparison of perturbative QCD calculations and experimental data. The details can be found in \cite{larkoski2014soft, kang2020recent}.~In another work \cite{ringer2020can} have studied jet-broadening in heavy ion collisions at EIC.~They have shown that jet grooming provides a new opportunity to investigate jet broadening effects.~This approach is orthogonal to the jet variables such as the azimuthal angle and jet-momentum used at HERA, which is more traditional.~Probing the same physics with independent observables offers an important cross-check to ensure consistency and predictive power of theoretical calculations.

~In the present work, we study the jets and jet-substructure in terms of the observables, the subjet multiplicity and differential jet shape in NC DIS events for different EIC energies.~These variables have been well studied at HERA, the first and only $ep$ collider, serves as a reference for future EIC jet measurements.~Such an analysis has not been investigated so far.~Also, we compare two Monte Carlo event generators for this analysis.~The results are useful for comparison with the data and measuring non-perturbative effects by further improving the event generator-tuning.~Subjet multiplicity is the number of jet-like substructures resolved within jets by reapplying the jet algorithm at a smaller resolution scale $y_{cut}$.~Jet shape is defined as the average fraction of the jet’s transverse energy $E_{T,jet}$ contained inside an annulus of radius $r$ \cite{zeus1999measurement}.

The subjet multiplicity distributions and their average values are studied for different jet sizes and resolution parameter.~Validity of $k_T$ and anti-$k_{T}$ algorithms for the production of jets and subjets in DIS $ep$ is also studied. Subjet multiplicity is studied as a function of number of jets $N_{jet}$, resolution parameter $y_{cut}$ and jet transverse energy $E_{T,jet}$.~Predictions for subjet multiplicity and differential jet shape are made for the upcoming EIC.
A detailed introduction to the subject of this paper is presented in Section I.~For the present study, kinematics of the neutral current deep inelastic scattering is discussed in Section-II. Details of data generation from different Monte Carlo event generators are given in Section-III.~Methods of jet and subjet formation and jet-shape study are detailed in Section-IV. Results are presented in Section-V, followed by conclusion in Section-VI.
\section{Kinematics of deep inelastic scattering}
\begin{figure}[ht]
\centering
\resizebox{6.5cm}{!}{\includegraphics{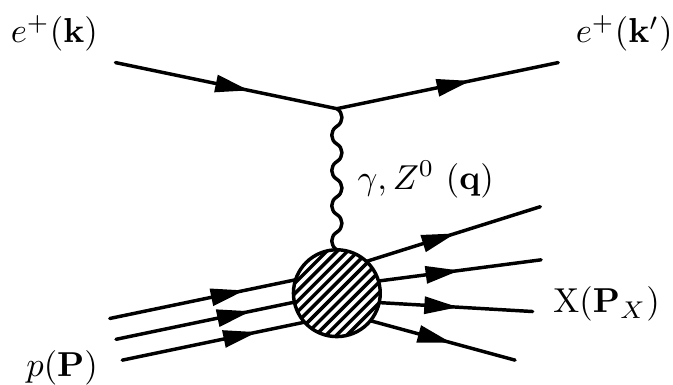}}
\caption{Neutral Current $e^{+}p$ Deep Inelastic Scattering}
\label{fig:dis}
\end{figure}
In deep inelastic scattering, a high-energy lepton  scatters off a hadron with large momentum transfer as shown in Figure \ref{fig:dis}.~For the present work, neutral current events are simulated which are mediated by the exchange of $\gamma$ or $Z$ boson.~Alternatively, if the scattering occurs via W boson then it is called charged current.~The kinematics of deep inelastic scattering can be described by the following Lorentz invariant variables \cite{devenish2011deep}.
The negative of the invariant mass squared of the exchanged virtual boson is represented by $Q^2$.~It can be interpreted as the power with which the exchanged boson can resolve the proton structure:
\begin{equation}
 Q^2 = - {\textbf{q}}.{\textbf{q}} = -(\textbf{k} - \textbf{k$'$} )^2 \\
\end{equation}
\begin{equation}
y = \textbf{P}.\textbf{q}/\textbf{P}.\textbf{k}
\end{equation}
where the inelasticity $y$, represents the fractional energy loss from lepton to the hadronic system.
Bjorken variable ($x$) is associated with the fraction of momentum of the proton carried by the struck parton:
\begin{equation}
 x = Q^2/2\textbf{P}.\textbf{q}
\end{equation}
The variables $Q^2$, $y$ and $x$ follow the relation:
\begin{equation}
Q^2 = sxy     
\end{equation}
 where $s = (\textbf{P} + \textbf{k})^2$ is square of the center of mass energy.
  W is the energy of the $\gamma^*$p system, which is equal to the total mass of the final state hadronic system X:
\begin{equation}
 W^2 = (\textbf{P} + \textbf{k})^2 \ \ (\equiv    \mathbf{P_{X}^2})
\end{equation}
\begin{figure}[ht!]
\centering
\resizebox{6cm}{!}{\includegraphics{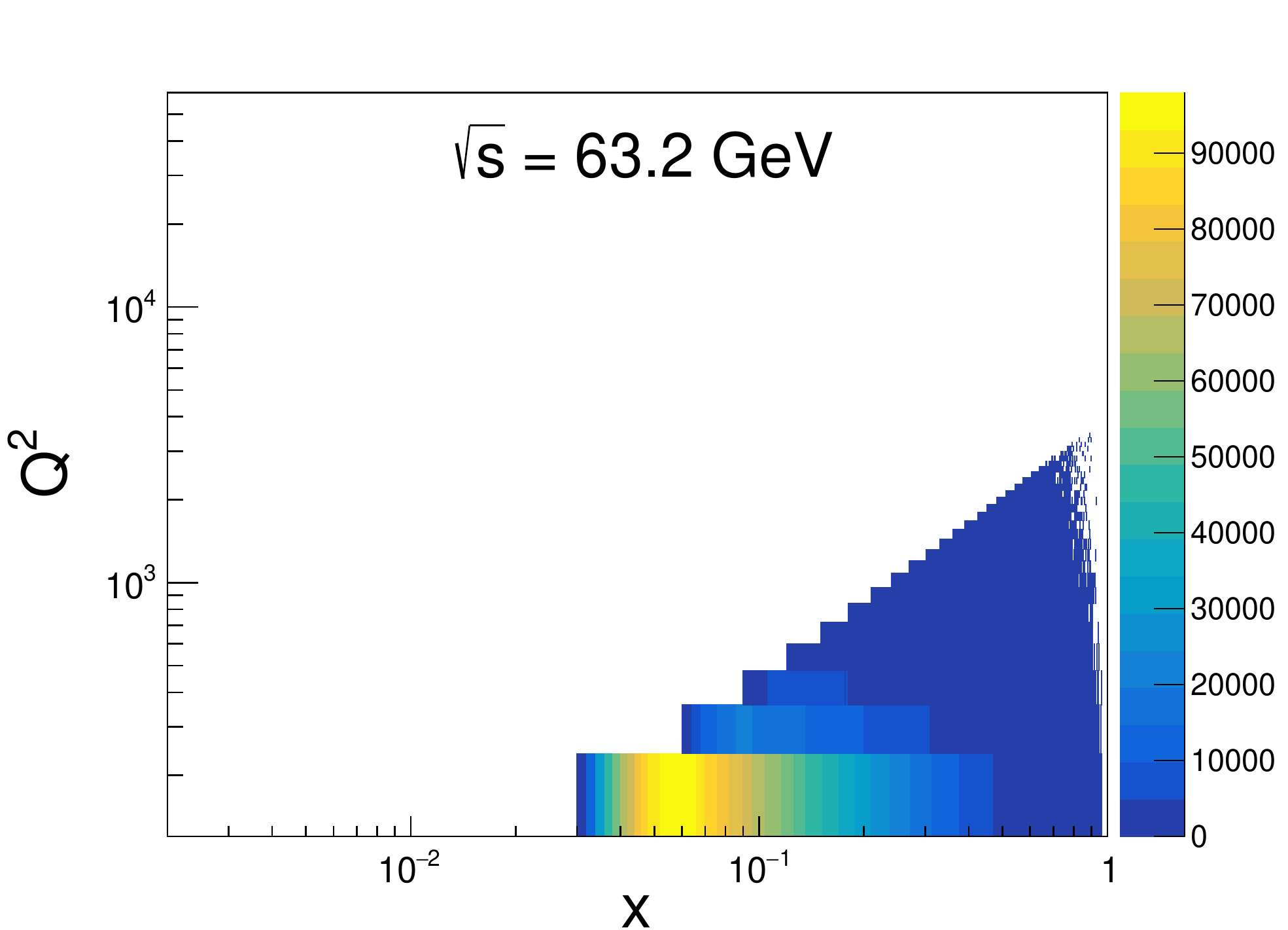}}
\label{scatter_eic1}
\end{figure}
\begin{figure}[ht!]
\centering
\resizebox{6cm}{!}{\includegraphics{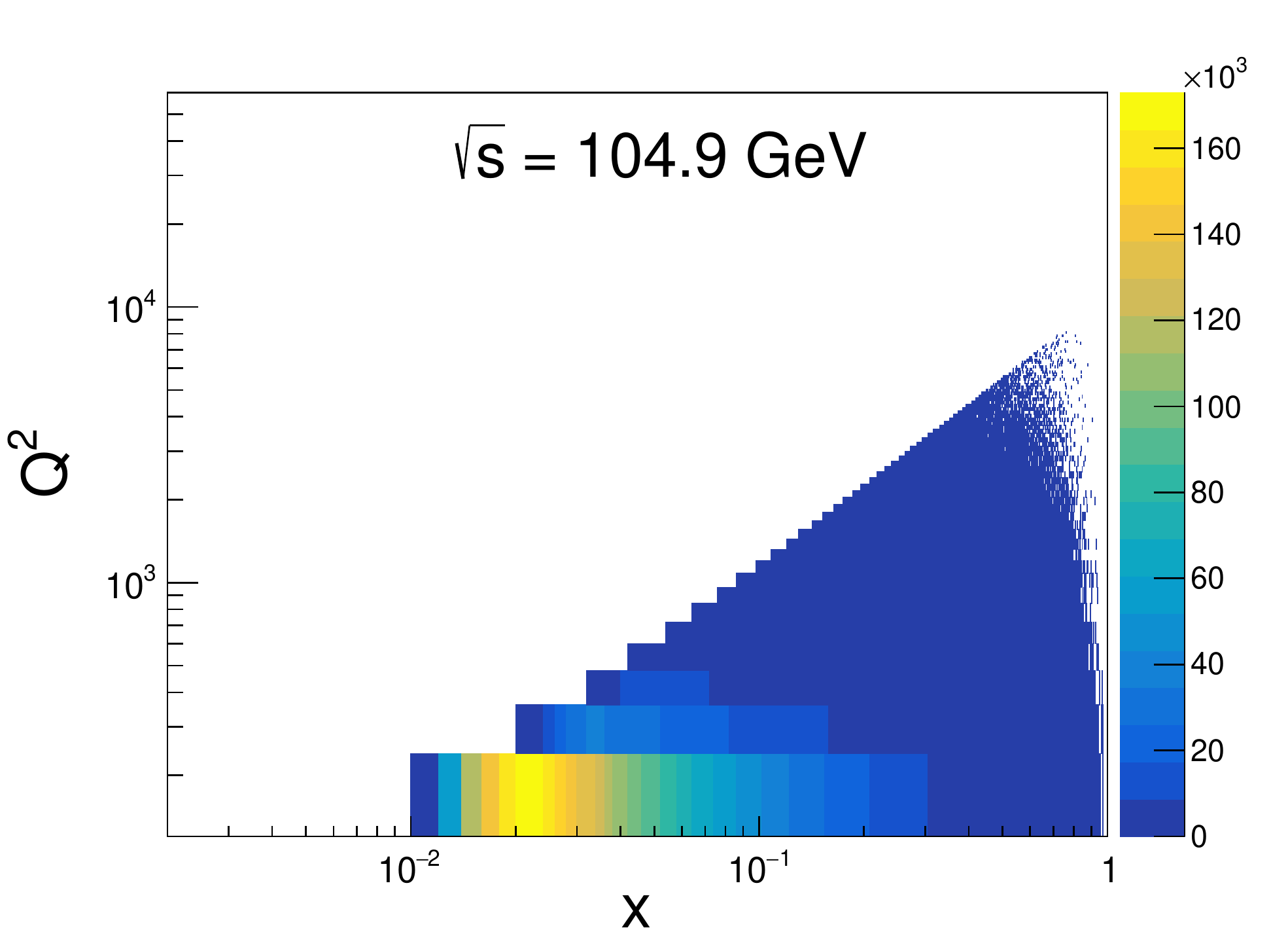}}
\label{scatter_eic2}
\end{figure}
\begin{figure}[ht!]
\centering
\resizebox{6cm}{!}{\includegraphics{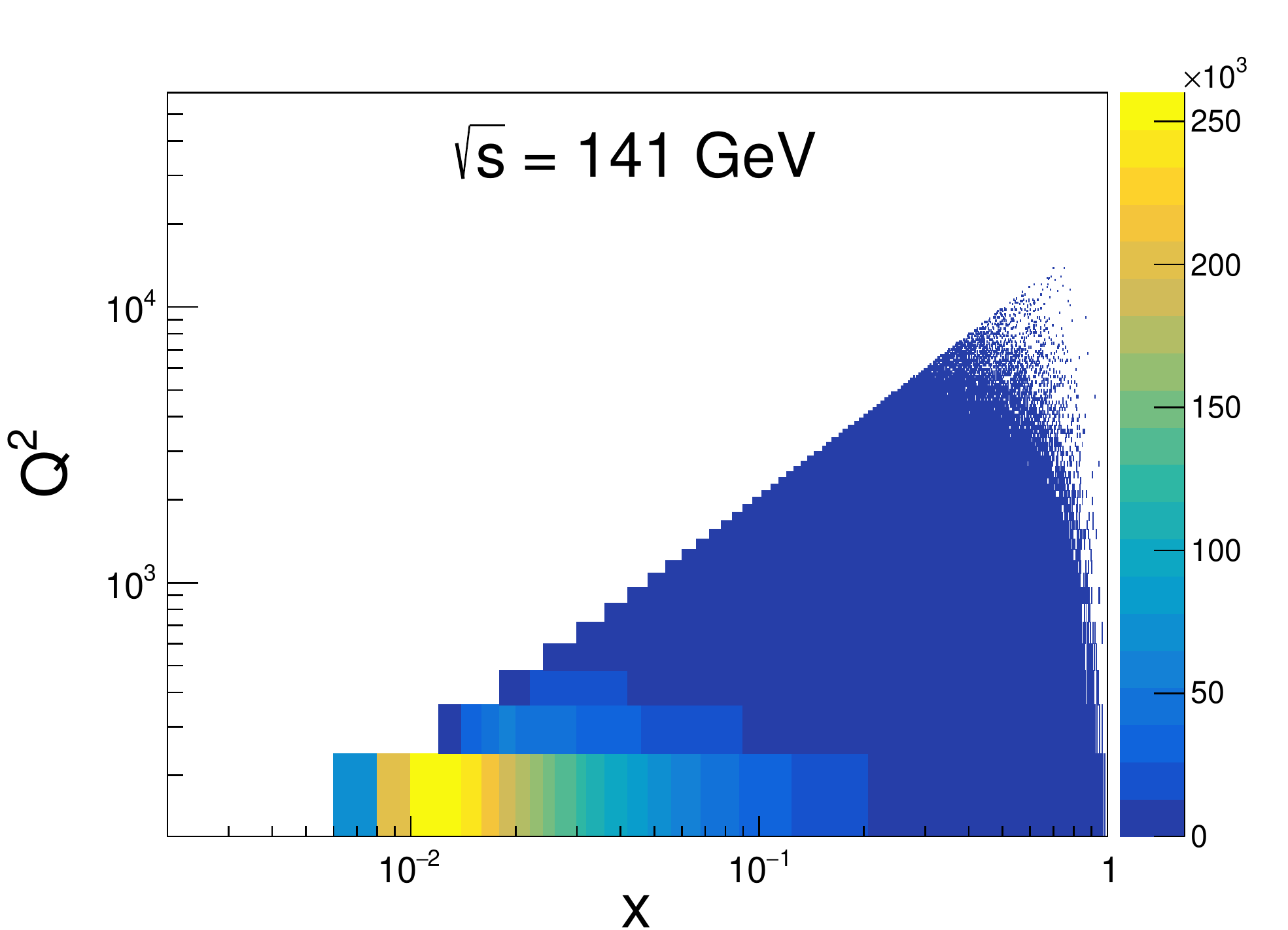}}
\label{scatter_eic3}
\end{figure}
\begin{figure}[ht!]
\centering
\resizebox{6cm}{!}{\includegraphics{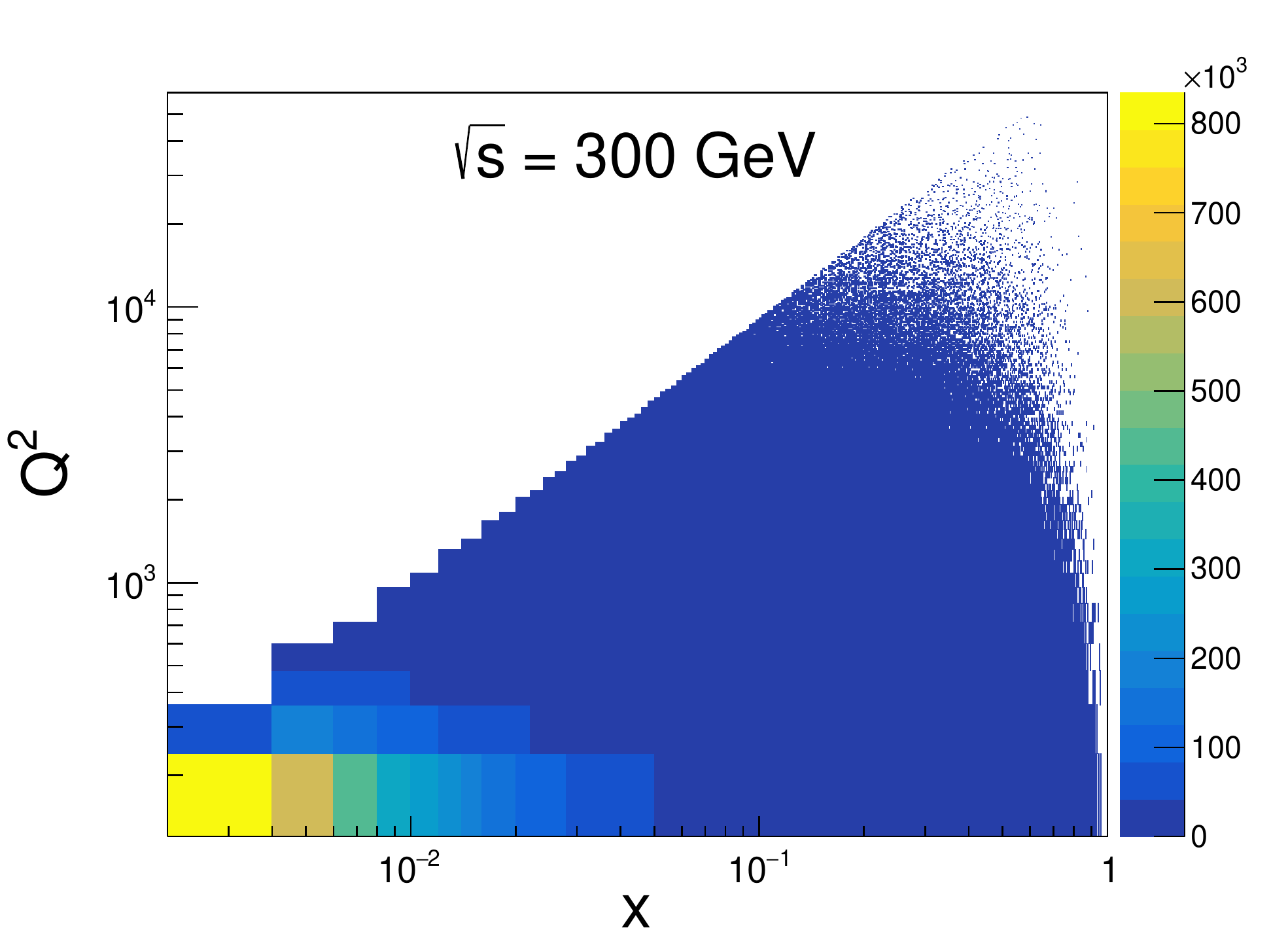}}
\caption{The range of Bjorken variable $x$ versus virtuality $Q^{2}$ for interactions at $\sqrt{s}$= 63.2, 104.9, 141 GeV at EIC and 300 GeV at HERA.}
\label{scatter_hera}
\end{figure}
\section{DATA GENERATION}
Monte Carlo event generators \cite{montecarlo} PYTHIA 8.304 \cite{pythia8} and RAPGAP 3.308 \cite{rapgapmain} are used to simulate $10^7$ NC DIS $e^+p$ events at different cms energies as shown in   
Table \ref{tab:my_label1}.~To validate the present work, we also simulated NC events at HERA energy of $\sqrt{s}$ = 300 GeV.~Jets are formed by implementing the $k_T$ sequential recombination algorithm with different jet radii R = 0.4, 0.6, 0.8 and 1.0  within the kinematic region $Q^2 > 125$ GeV$^2$, where $Q^2$ is the virtuality of the exchanged boson and jets are identified with $E_{T jet} > 10$ GeV. Subjets are resolved by re-running the $k_T$ algorithm on the jets and using a resolution parameter, $y_{cut}$.
For both the event generators, QCD cascade is taken to be $p_{t}$ ordered parton showers with all QED radiations turned off. The parton distribution function (PDF) set NNPDF 2.3 \cite{nnpdf} with $\alpha_{s}$ = 0.130 is implemented in both the generators through LHAPDF package 6.4.0 \cite{lhapdf}. PYTHIA 6.4 \cite{pythia6main} is used for fragmentation in RAPGAP. For jet finding and implementation of sequential recombination clustering algorithms, FastJet 3.4.0 \cite{fastjet} package is used on data samples from each event generator. 
\begin{table}[h!]
    \centering
    \scalebox{1.3}{
    \scriptsize
    \begin{tabular}{|c|c|c|c|}
    \hline
    Positron & Proton & $\sqrt{s}$ (GeV) & Collider\\
    energy (GeV) & energy (GeV) & & \\
\hline
 10 & 100 & 63.2 & EIC\\
\hline
 10 & 275 & 104.9 & EIC\\
\hline
 20 & 250 & 141 & EIC \\
 \hline
  27.5 & 820 & 300 & HERA\\
\hline
    \end{tabular}}
    \caption{The beam energies for colliding particles and the center of mass energies at EIC and HERA colliders. }
    \label{tab:my_label1}
\end{table}
Figure \ref{scatter_hera} shows the range of Bjorken variable $x$ versus virtuality $Q^{2}$ for center of mass energies at EIC and HERA.~EIC aims to provide an enhanced range to investigate quarks and gluons with small momentum fraction ($x$) and analyse their properties over a wide range of momentum transfers $Q^{2}$ \cite{EIC2}. 
\section{JETS}
Quarks and gluons are the fundamental entities produced in high-energy interactions, which instantly fragment and hadronize producing collimated sprays of energetic hadrons.~An algorithm is used to cluster these hadrons into jets and several clustering algorithms have been developed.~A review of the jet algorithms can be found in the paper \cite{salam2010towards}.~Number of jet forming algorithms \cite{salam2010towards} are available with $k_{T}$ \cite{ktalgorithm,ktalgorithm2} and anti-$k_T$ \cite{antiktalgorithm} being the most utilised to identify the jets and subjets.
\begin{figure}[ht]
\centering
\resizebox{8cm}{!}{\includegraphics{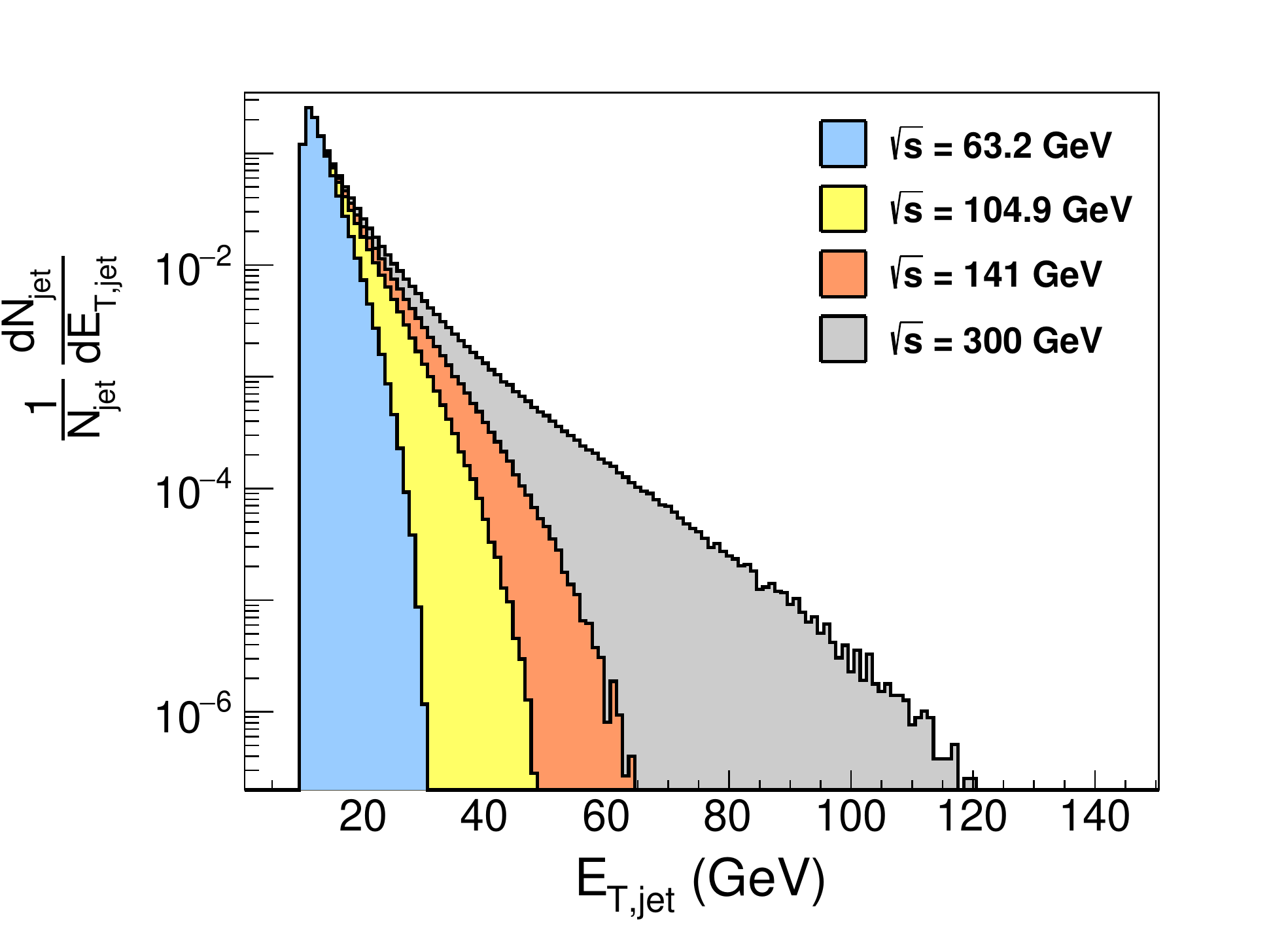}}
\caption{Jet transverse energy $E_{T,jet}$ for each center of mass energy with R = 1.}
\label{Et_jet}
\end{figure}
\begin{figure}[ht]
\centering
\resizebox{8cm}{!}{\includegraphics{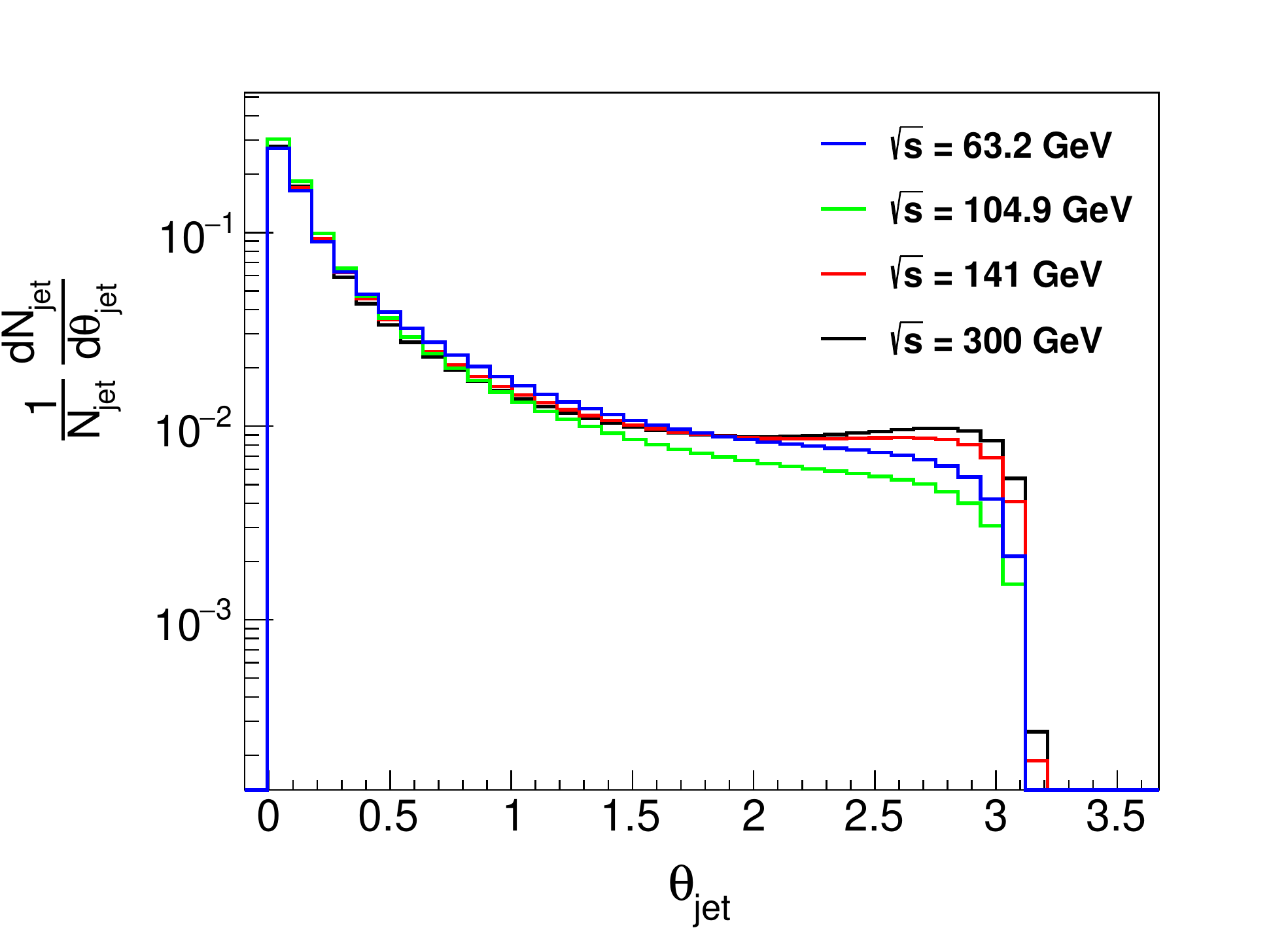}}
\caption{Jet angle $\theta_{jet}$ defined in equation~(8), for jets with R=1 at different $\sqrt{s}$}
\label{eta_jet}
\end{figure}
Sequential clustering algorithms, particularly $k_{T}$ and anti-$k_{T}$ algorithms, are used to cluster the final-state hadrons for the production of jets.~Subjets are obtained by reapplying the jet algorithms. The procedure for generalised $k_{T}$ jet algorithm is:
\begin{itemize}
\item For every pair of $i^{th}$ and $j^{th}$ particles, inter-particle and particle-beam distance is defined as:
\begin{gather} \label{jet_algorithm}
d_{ij} = min(p^{2\rho}_{t_i}, p^{2\rho}_{t_j})\frac{(y_{i} - y_{j})^2 + (\phi_{i} - \phi_{j})^2}{R^2}\\
d_{iB} = p_{ti}^{2\rho}
\end{gather}
$p_{t_i}$, $y_{i}$ and $\phi_{i}$ are the transverse momentum with respect to the beam direction, rapidity, and azimuth respectively of the particle $i$. R is a free parameter usually called the jet radius.
\item The smallest distance among all the $d_{ij}$ and $d_{iB}$ are found iteratively.~If the minimum distance is $d_{ij}$, then the $i^{th}$ and $j^{th}$ particles are merged into single cluster and removed from the list.~And if it is $d_{iB}$, then $i^{th}$ particle is called a jet and removed from the list.
\end{itemize}
Generalized $k_{T}$ algorithm reduces to $k_{T}$ jet algorithm for $\rho = 1$, Cambridge/Aachen jet algorithm \cite{cambridge} for $\rho = 0$, and anti-$k_{T}$ jet algorithm for $\rho = -1$. All of these algorithms are infrared and collinear (IRC) safe. By equation \ref{jet_algorithm}, $k_{T}$ algorithm \cite{ktalgorithm,ktalgorithm2} is dominated by low $p_{t}$ and clusters soft particles first which results in a jet area that differs significantly and makes it more susceptible to underlying events (UE) and pile-up (PU).~Whereas, anti-$k_{T}$ algorithm \cite{antiktalgorithm} prefers to cluster hard particles first, and the jet area only differs marginally making the algorithm only slightly susceptible to the UE and PU \cite{atkin2015review}.~In general any algorithm can produce jets with one particle, but the implementation of kinematic selections minimises such events.~Figures \ref{Et_jet} and \ref{eta_jet} show the distributions of jet transverse energy $E_{T,jet}$ and jet angle $\theta_{jet}$ respectively for jets produced by $k_T$ algorithm at each energy.~$\theta_{jet}$ is measured from jet pseudorapidity $\eta_{jet}$ as follows:
\begin{equation}
    \theta_{jet} = 2\arctan{(e^{-\eta_{jet}})}
\end{equation}
\subsection{Subjets}
\begin{figure}[ht]
\centering
\resizebox{8.5cm}{!}{\includegraphics{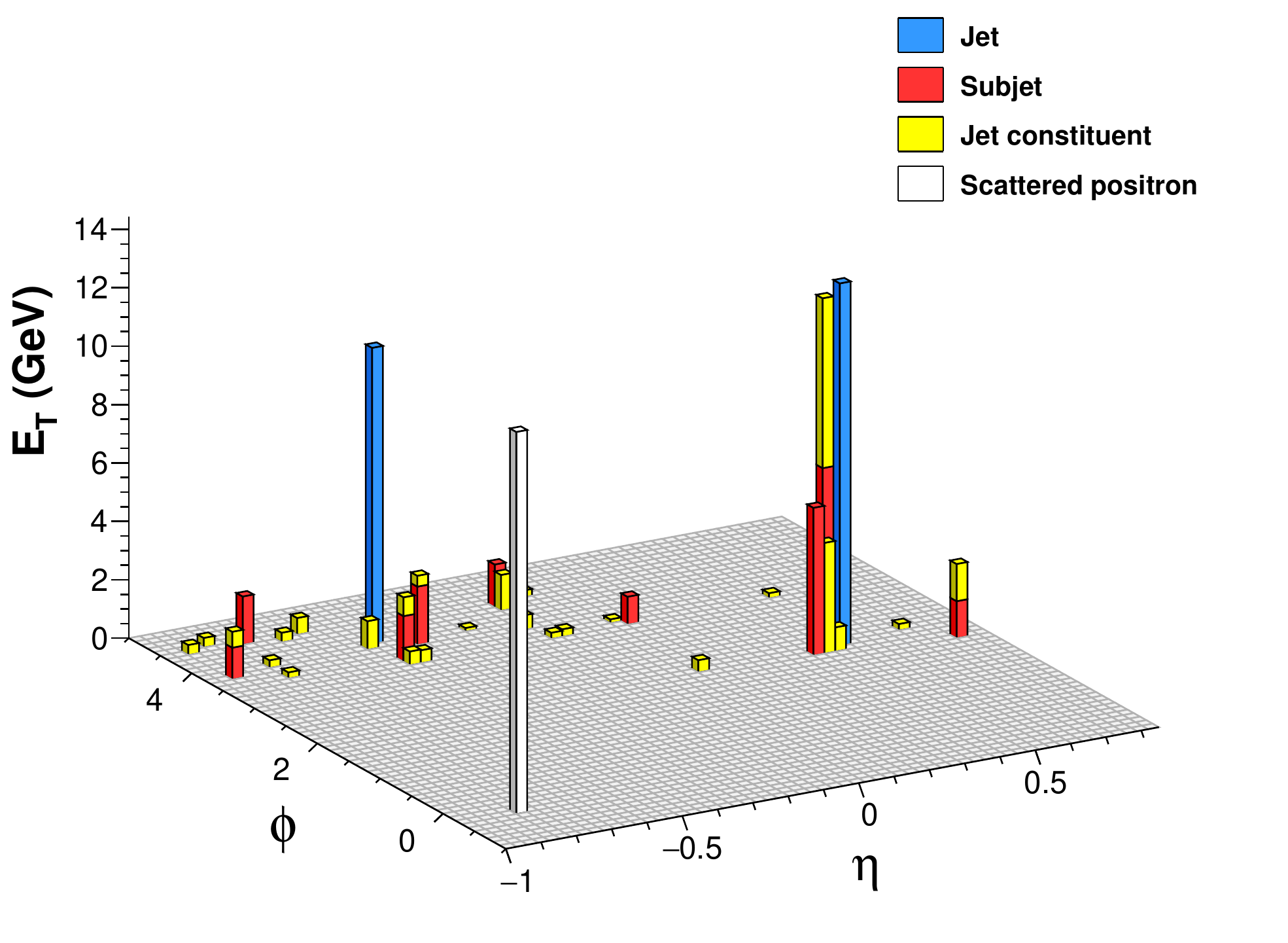}}
\caption{An example di-jet event with constituents, subjets and scattered positron simulated from PYTHIA 8.304.~The number of subjets are three and seven in respective jets.}
\label{3dplot}
\end{figure}

Subjets are used to probe the internal structure of jets, and are produced by reapplying procedure for jet algorithm on constituents of jets as input particles. The procedure from jet algorithm is repeated until all particles satisfy the following condition:
\begin{equation}
d_{cut} = y_{cut}(E_{T,jet}^2).
\end{equation}
Figure \ref{3dplot} shows an example di-jet event with constituents, subjets and scattered positron simulated by PYTHIA 8.304.~The subjet multiplicity $({n_{sbj}})$ is dependent on the selected values of resolution parameter $y_{cut}$.~The average number of subjets resolved within a jet at a specific $y_{cut}$ value is defined as mean subjet multiplicity $\langle n_{sbj}\rangle$:
The perturbative QCD value of $\langle n_{sbj}\rangle$ can be calculated as the ratio of the cross section for subjet production to that for inclusive jet production ($\sigma_{jet}$).
\begin{equation}
\langle n_{sbj(y_{cut})}\rangle = 1 + \frac{1}{\sigma_{jet}}\sum\limits^{\infty}_{j = 2}(j - 1)\sigma_{sbj,j}(y_{cut}),
\end{equation}
where, $\sigma_{sbj,j}(y_{cut})$ is the cross section for producing jets with $j$ subjets at a resolution scale of $y_{cut}$.
The mean subjet multiplicity $\langle n_{sbj}(y_{cut})\rangle$ is measured by \cite{zeusmain}:
\begin{equation}
\langle n_{sbj}(y_{cut})\rangle =\frac{1}{N_{jets}}\sum\limits^{N_{jets}}_{i = 1} n_{sbj}^{i}(y_{cut}),   
\end{equation}
where, $N_{jets}$ is the total number of jets in the sample and $n^i_{sbj}(y_{cut})$ are total resolved subjets in the $i^{th}$ jet.
\subsection{Differential jet shape}
\begin{figure}[ht]
\centering
\resizebox{6.5cm}{!}{\includegraphics{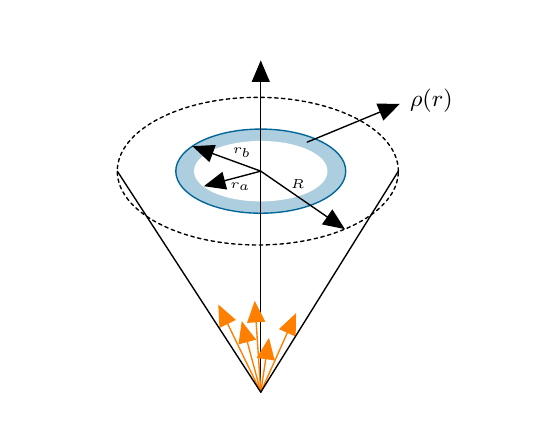}}
\caption{The differential jet shape is the average fraction of the jet’s transverse energy $E_{T,jet}$ contained inside an annulus of radius r, centered around the jet-axis.}
\label{fig:shapejet}
\end{figure}

The differential jet shape is defined as the average fraction of the jet’s transverse energy $E_{T,jet}$ contained inside an annulus of inner radius $r_{a}$($r-\Delta{r}/2$) and outer radius $r_{b}$ ($r+\Delta{r}/2$) concentric to the jet-axis in $\eta$ - $\phi$ plane.
\begin{equation}\label{equation_jetshape}
 \rho(r) = \frac{1}{N_{jets}}\frac{1}{\Delta{r}}\sum\limits_{jets} \frac{E_{T}(r-\Delta{r}/2,r+\Delta{r}/2)}{E_{T}(0,R)}
\end{equation}

where $N_{jets}$ is the total number of jets with jet radius R = 1.~Differential jet shape is studied as a function of distance $r = \sqrt{(\Delta{\eta})^2 + (\Delta{\phi})^2}$ from the jet-axis, with $\Delta{r} = 0.1$ increments. Figure \ref{fig:shapejet} shows the representation of differential jet shape at any radius $r$.

\section{Results and discussion}
Samples of $10^{7}$ NC DIS $e^{+}p$ collisions are simulated by two Monte Carlo Event generators PYTHIA 8.304 and RAPGAP 3.308 for center of mass energies $\sqrt{s}$ = 63.2, 104.9, 141 GeV at the EIC and $\sqrt{s}$ = 300 GeV at HERA for comparison.~The kinematic region is defined by $Q^2 >$ 125 GeV$^2$.~Final state hadrons in each event are clustered by using longitudinally invariant $k_{T}$ jet algorithm for different jet radii.~Only the jets with transverse energy $E_{T,jet} > $10 GeV are studied.~Jet finding and algorithm implementation is done using FastJet package \cite{fastjet}.~For each cms energy, one-jet events are found to be dominant in NC DIS $e^{+}p$.~The $E_{T,jet}$ distributions in Figure \ref{Et_jet} show that at $\sqrt{s} = 300$ GeV, jets with the highest $E_{T,jet}$ can be found up to 130 GeV.~While, for $\sqrt{s}$ = 63.2, 104.9 and 141 GeV, the highest $E_{T,jet}$ is observed to decrease with values around 30, 50 and 60 GeV respectively.~It is also observed that most of the jets are produced in the forward direction as shown in Figure \ref{eta_jet}, making very small angles to the initial hadron beam direction.~They can however, be produced significantly in the barrel region.
\subsection{Subjet Multiplicity}
\begin{figure}[ht]
\centering
{\resizebox{6cm}{!}{\includegraphics{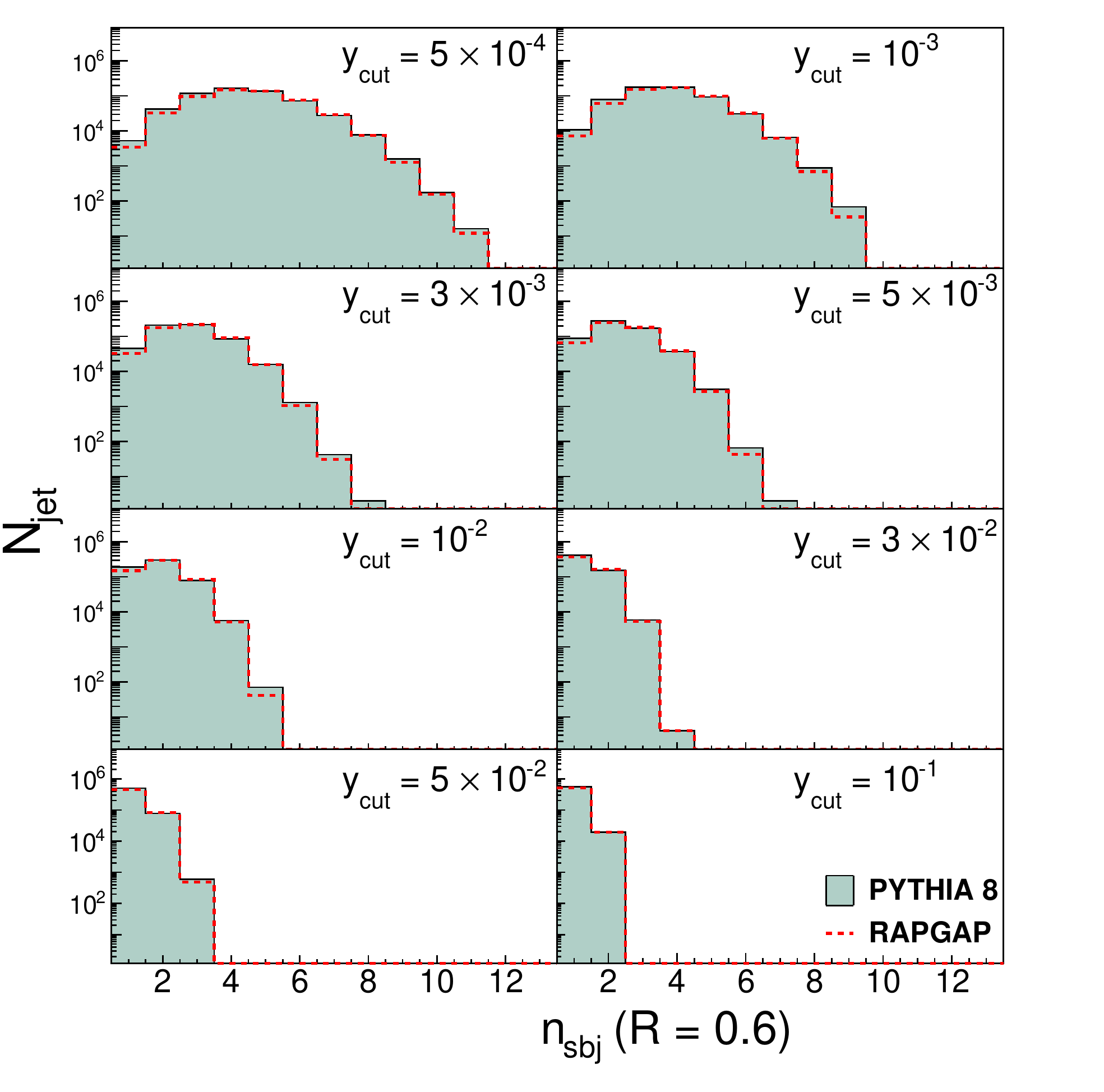}}
\resizebox{6cm}{!}{\includegraphics{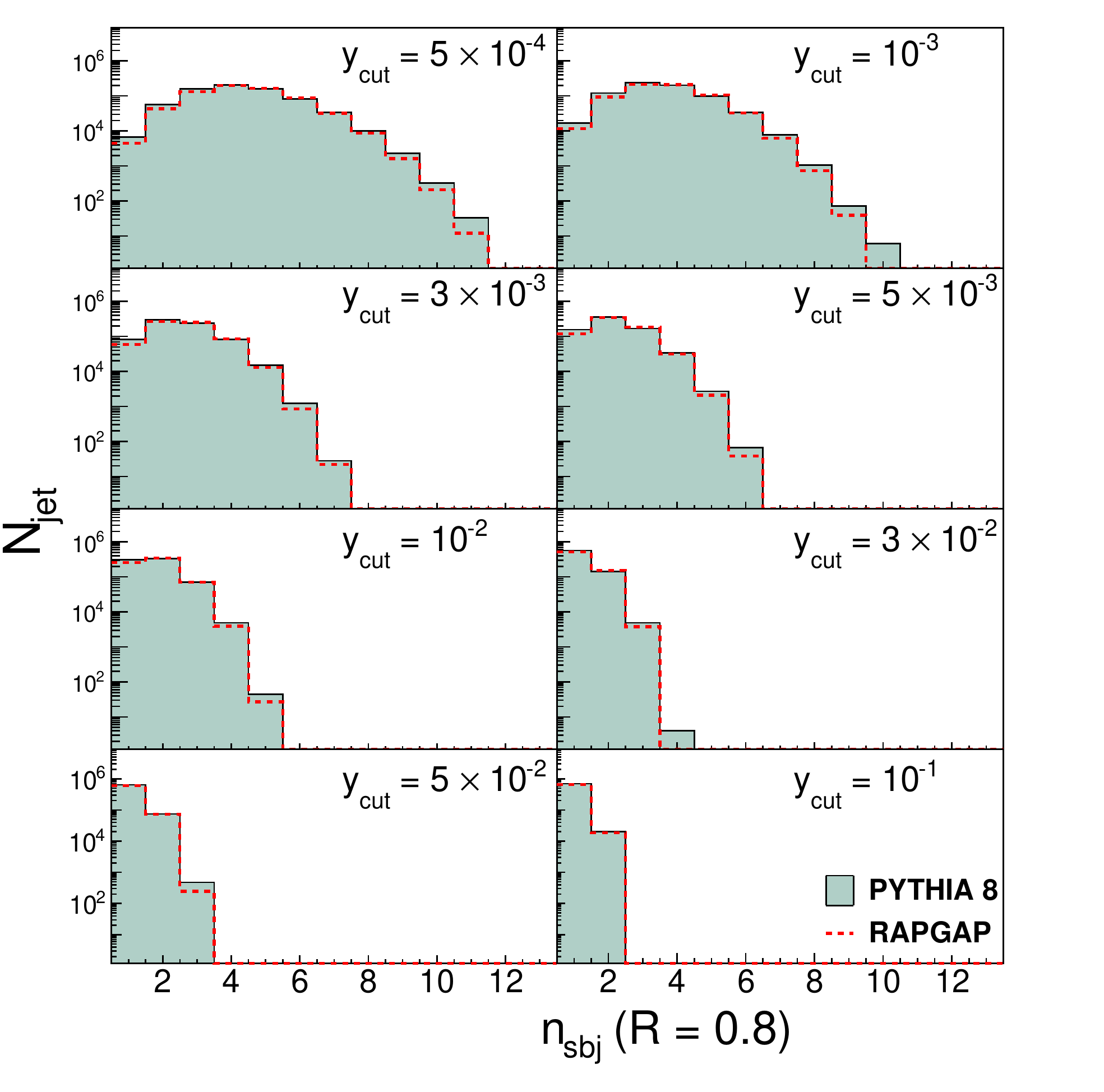}}
\resizebox{6cm}{!}{\includegraphics{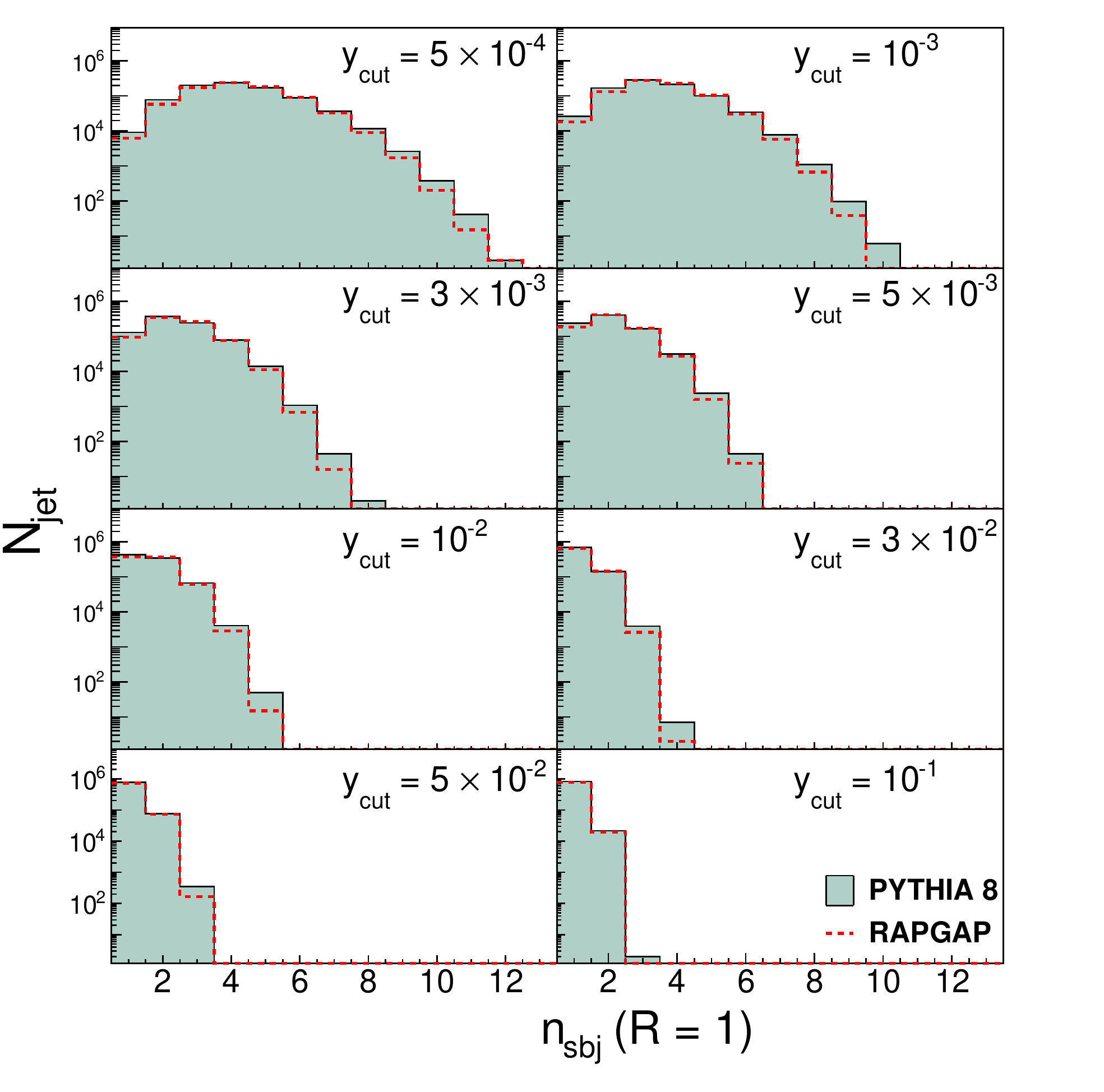}}}
\caption{Subjet multiplicity for EIC at $\sqrt{s}$ = 63.2 GeV for R = 0.6, 0.8 and 1 from top to bottom.}
\label{06_nsbj_eic1}
\end{figure}

\begin{figure}[ht]
\centering
{\resizebox{6cm}{!}{\includegraphics{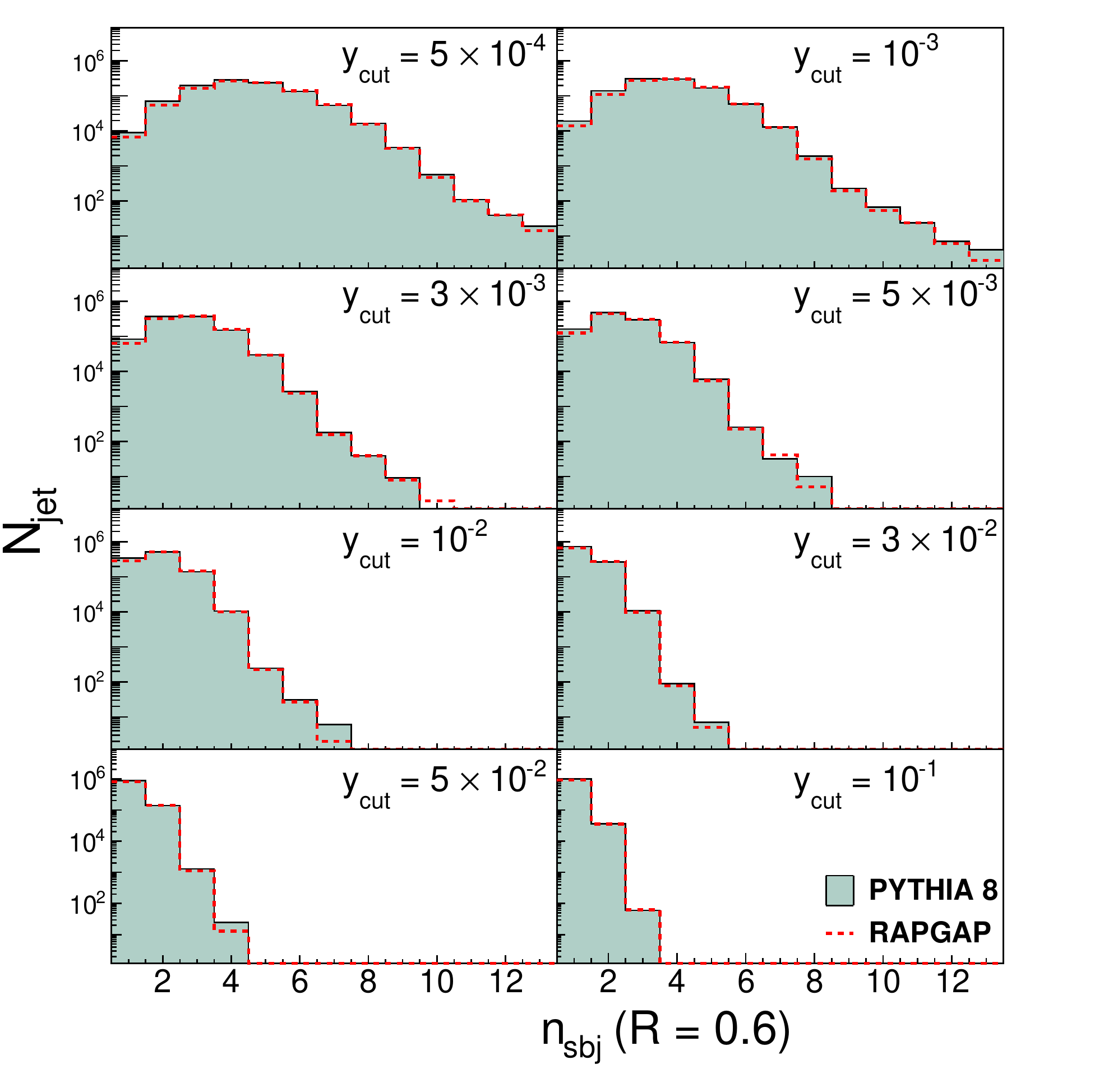}}
\resizebox{6cm}{!}{\includegraphics{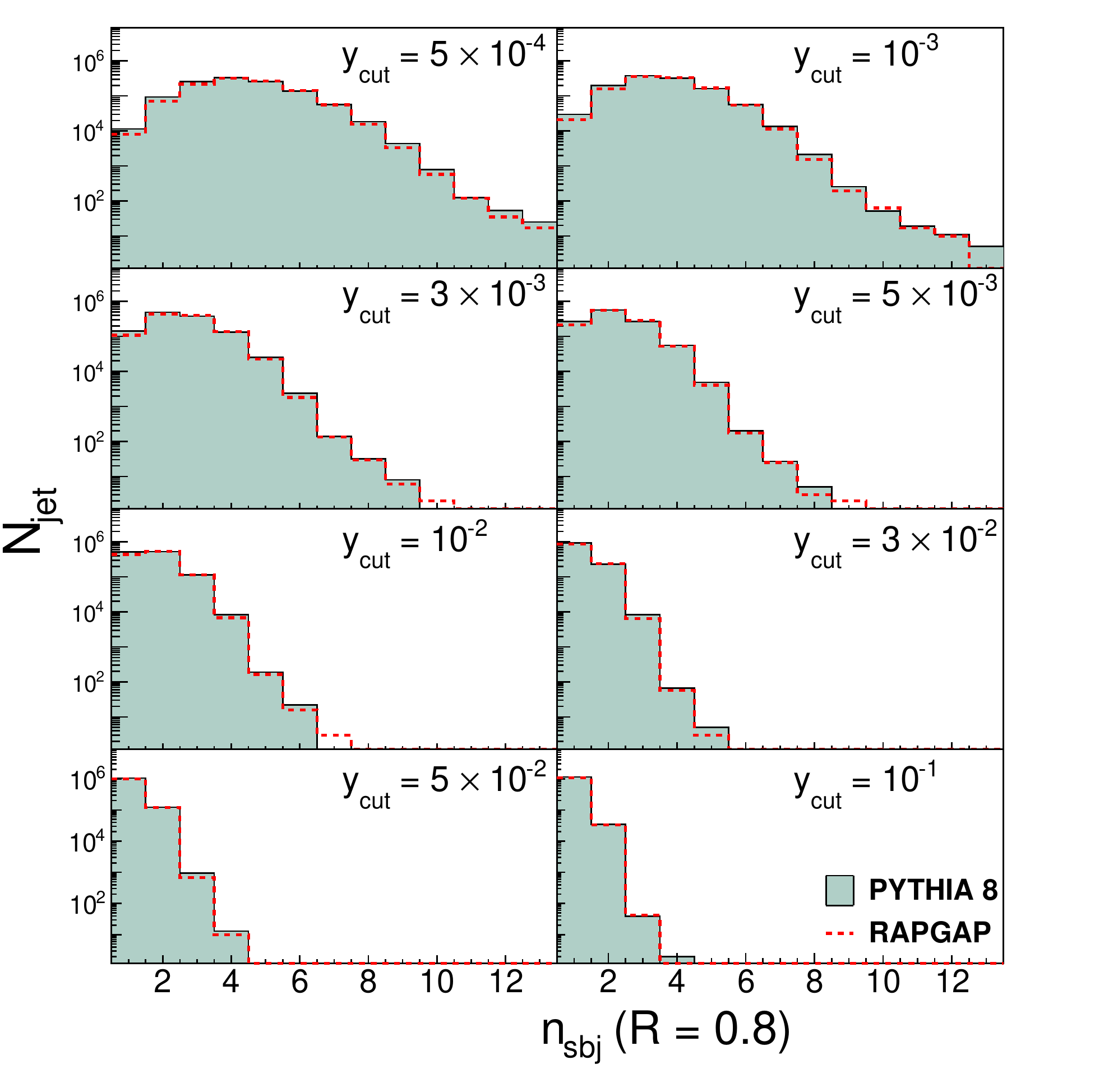}}
\resizebox{6cm}{!}{\includegraphics{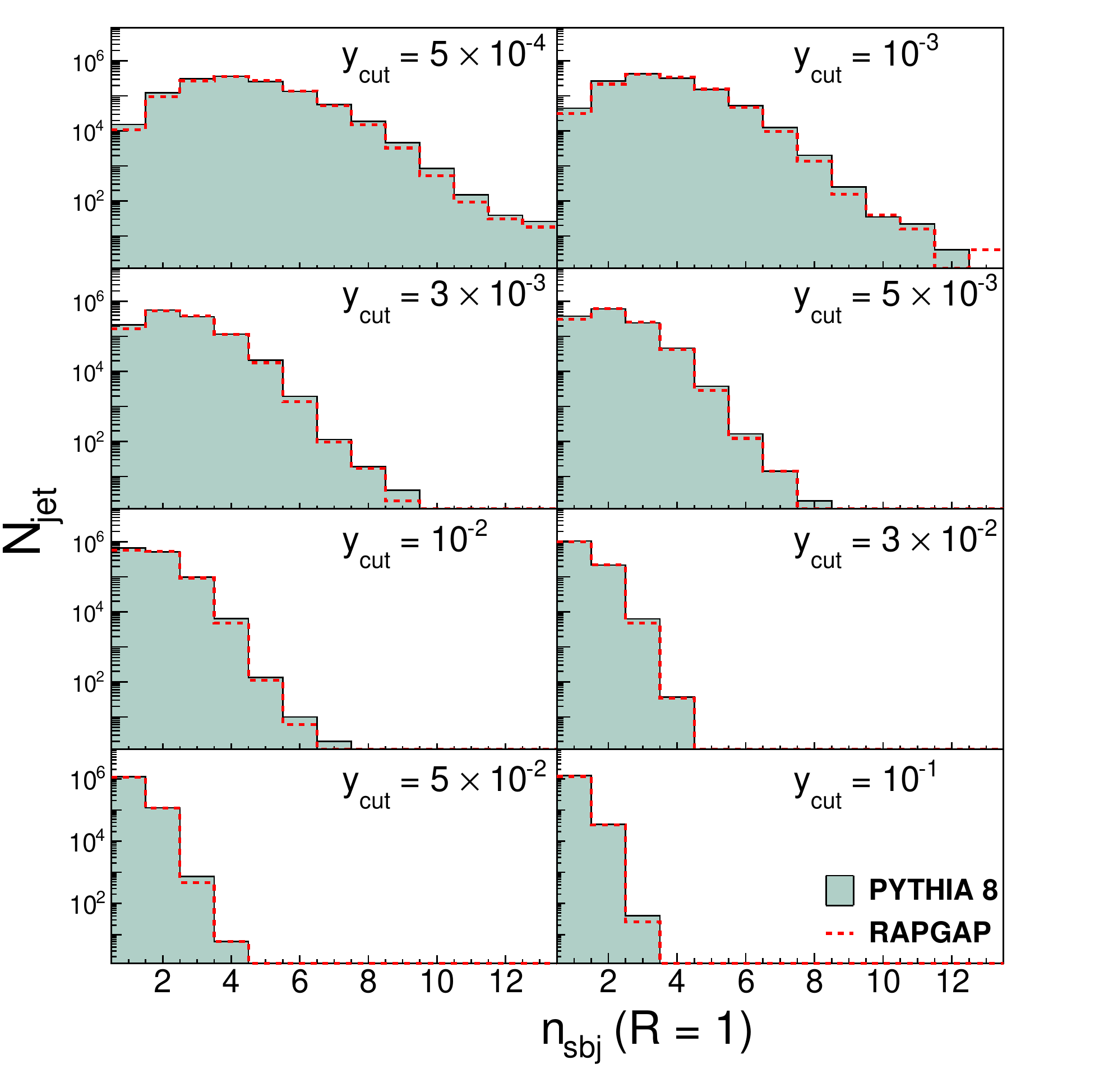}}}
\caption{Subjet multiplicity for EIC at $\sqrt{s}$ = 104.9 GeV for R = 0.6, 0.8 and 1 from top to bottom.}
\label{06_nsbj_eic2} 
\end{figure}

\begin{figure}[ht]
\centering
{\resizebox{6cm}{!}{\includegraphics{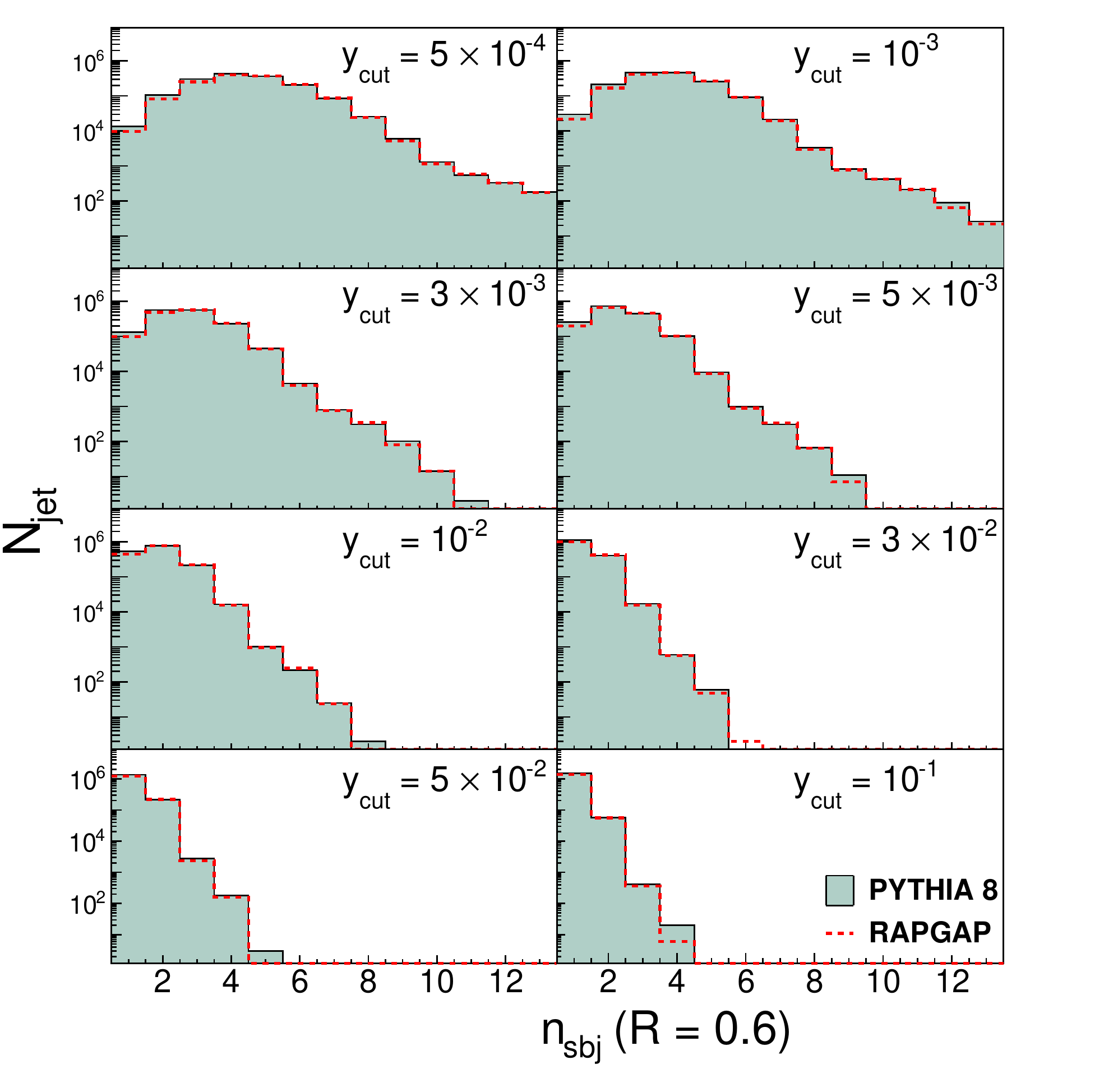}}
\resizebox{6cm}{!}{\includegraphics{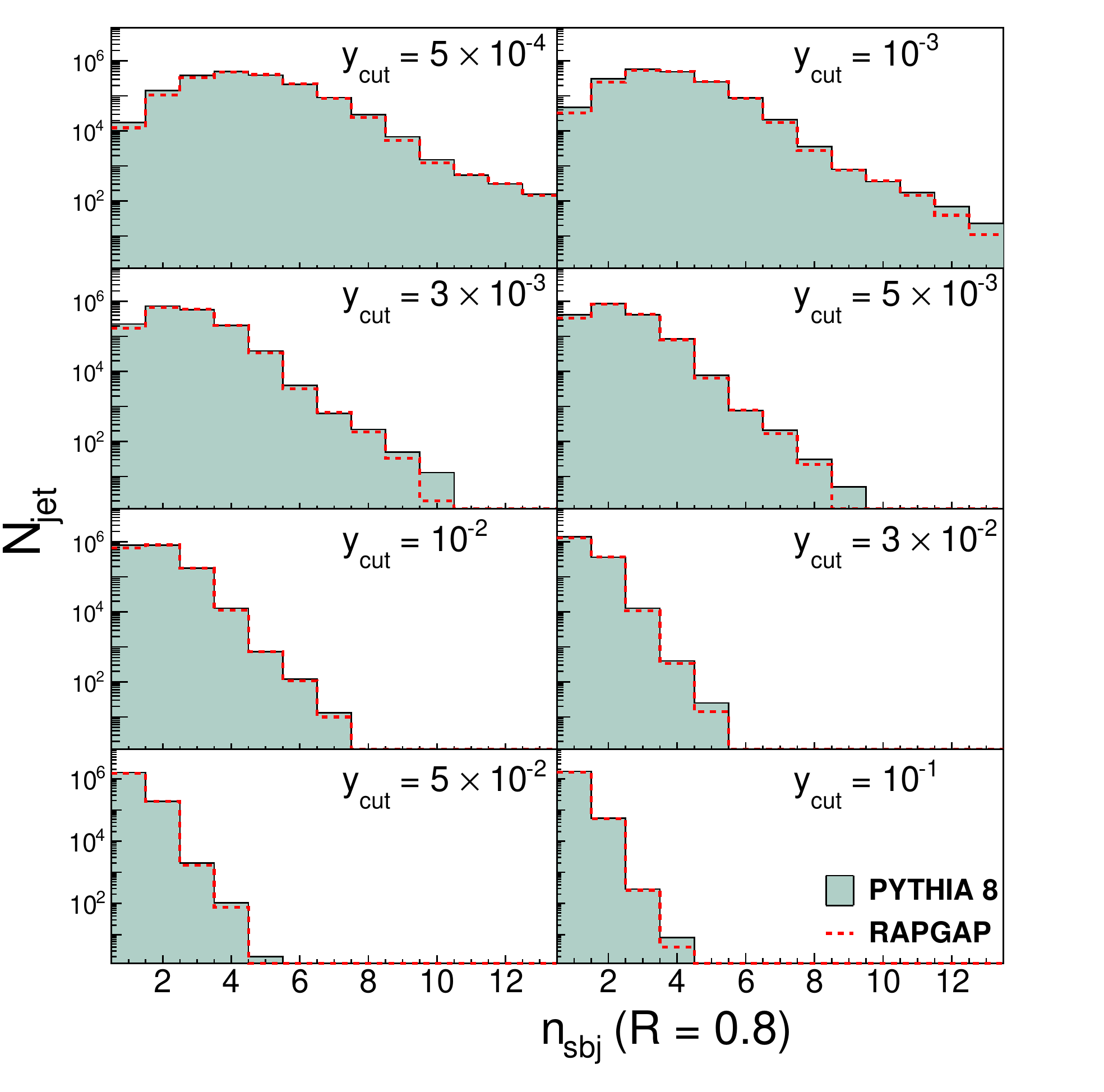}}
\resizebox{6cm}{!}{\includegraphics{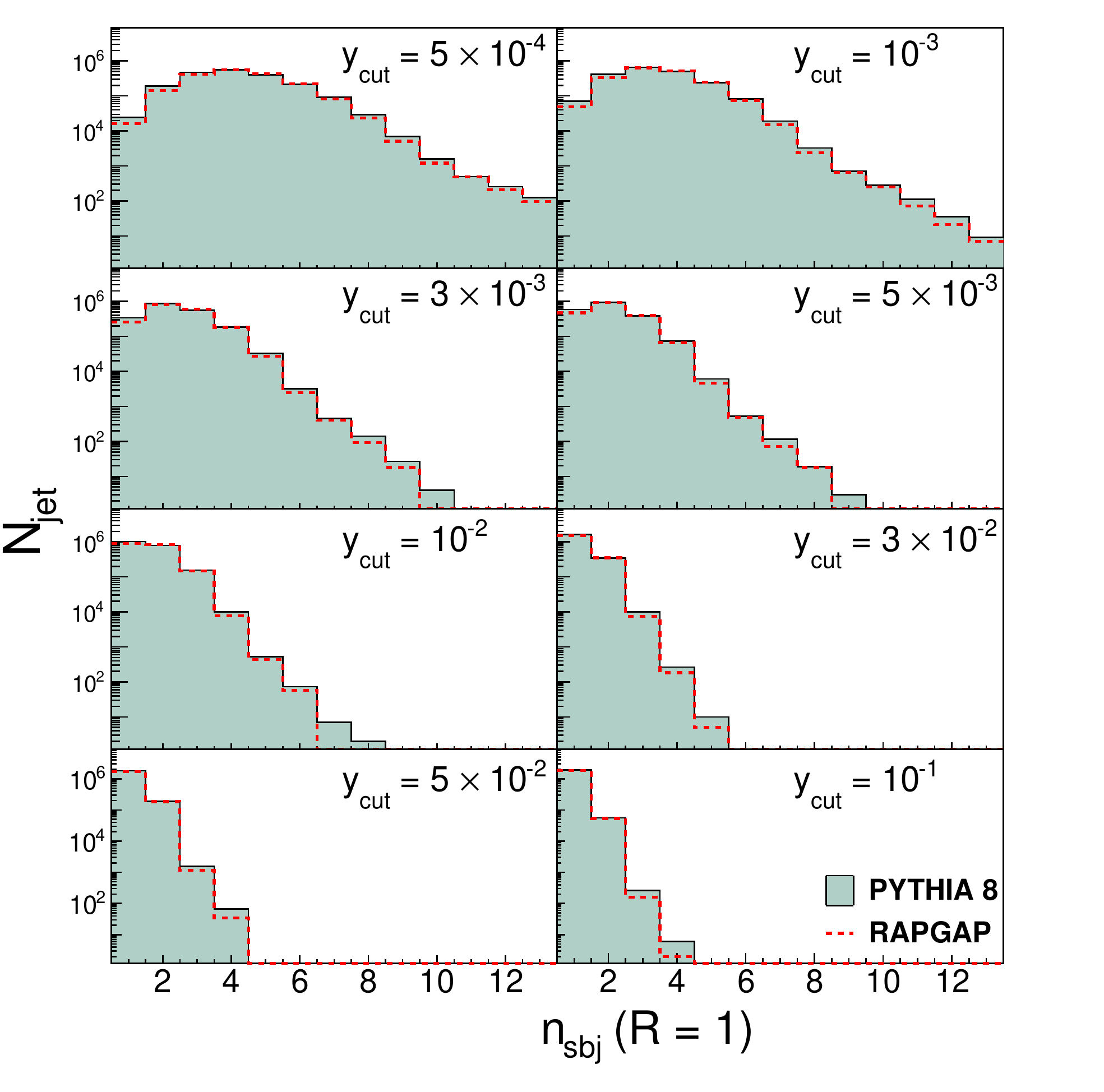}}}
\caption{Subjet multiplicity for EIC at $\sqrt{s}$ = 141 GeV for R = 0.6, 0.8 and 1 from top to bottom.}
\label{06_nsbj_eic3} 
\end{figure}

Subjets are formed by reapplying longitudinally invariant $k_{T}$ jet algorithm on hadronic jets at smaller resolution parameter $y_{cut}$.~Events with jet transverse energy $E_{T,jet} >$ 15 GeV and jet pseudorapidity -1 $< \eta_{jet} <$ 2 are used for forming the subjets.~The subjet multiplicity measurements at $\sqrt{s}$ = 63.2, 104.9 and 141 GeV, at various resolution parameter $y_{cut}$ are shown in Figures \ref{06_nsbj_eic1}, \ref{06_nsbj_eic2}, \ref{06_nsbj_eic3} for radius 0.6, 0.8 and 1 respectively.
It is observed that the subjet multiplicity decreases as $y_{cut}$ increases from 0.0005 to 0.1 for all jet radii.~Table \ref{tab:1} shows mean subjet multiplicity for each value of $y_{cut}$ and jet radius R for both event generators. It can also be seen that for a given jet radius R, $y_{cut}$ and minimum  $E_{T,jet}$ values of subjet multiplicities are similar for each center of mass energy, inferring that production of subjets in a jet of given transverse energy is independent of center of mass energy. RAPGAP gives slightly higher subjet multiplicity than PYTHIA.~For each generator, and for each $y_{cut}$ value, the subjet multiplicity decreases very marginally for increasing R.~However for the largest $y_{cut}$ of 0.1, it becomes roughly constant.

\begin{table}[h!]
    \centering
    \scalebox{0.95}{
    \scriptsize
    \begin{tabular}{|c|c|c|c|c|c|c|}
    \hline
     &\multicolumn{6}{c|}{$\langle n_{sbj} \rangle$ for $\sqrt{s} = 63.2$ GeV}\\
    \cline{2-7}
     {\large $y_{cut}$}  & \multicolumn{2}{c|}{R = 0.6} & \multicolumn{2}{c|}{R = 0.8} & \multicolumn{2}{c|}{R = 1}\\
     \cline{2-7}
       & PYTHIA & RAPGAP & PYTHIA & RAPGAP & PYTHIA & RAPGAP\\
 \hline 
 0.0005 & 4.32 & 4.46 & 4.26 & 4.37 & 4.16 & 4.25 \\
 \hline 
 0.001 & 3.67 & 3.78 & 3.53 & 3.64 & 3.40 & 3.49 \\
 \hline 
 0.003 & 2.69 & 2.79 & 2.51 & 2.60 & 2.38 & 2.45 \\
 \hline 
 0.005 & 2.29 & 2.38 & 2.12 & 2.20 & 2.00 & 2.06 \\
 \hline 
 0.01 & 1.82 & 1.90 & 1.68 & 1.74 & 1.58 & 1.63 \\
 \hline 
 0.03 & 1.28 & 1.32 & 1.21 & 1.23 & 1.18 & 1.19 \\
 \hline 
 0.05 & 1.14 & 1.16 & 1.10 & 1.11 & 1.09 & 1.09 \\
 \hline 
 0.1 & 1.03 & 1.04 & 1.03 & 1.03 & 1.03 & 1.02 \\
 \hline 
 
       &\multicolumn{6}{c|}{$\langle n_{sbj} \rangle$  for $\sqrt{s} = 104.9$ GeV}\\
    \cline{2-7}
    {\large $y_{cut}$} & \multicolumn{2}{c|}{R = 0.6} & \multicolumn{2}{c|}{R = 0.8} & \multicolumn{2}{c|}{R = 1}\\
     \cline{2-7}
       & PYTHIA & RAPGAP & PYTHIA & RAPGAP & PYTHIA & RAPGAP\\
 \hline 
 0.0005 & 4.39 & 4.50 & 4.28 & 4.38 & 4.15 & 4.24 \\
 \hline 
 0.001 & 3.70 & 3.79 & 3.53 & 3.62 & 3.38 & 3.46 \\
 \hline 
 0.003 & 2.69 & 2.77 & 2.50 & 2.58 & 2.35 & 2.42 \\
 \hline 
 0.005 & 2.28 & 2.36 & 2.11 & 2.17 & 1.98 & 2.03 \\
 \hline 
 0.01 & 1.82 & 1.88 & 1.67 & 1.72 & 1.57 & 1.61 \\
 \hline 
 0.03 & 1.28 & 1.31 & 1.21 & 1.23 & 1.18 & 1.19 \\
 \hline 
 0.05 & 1.14 & 1.15 & 1.11 & 1.11 & 1.09 & 1.09 \\
 \hline 
 0.1 & 1.04 & 1.04 & 1.03 & 1.03 & 1.03 & 1.03 \\
 \hline 
 
       &\multicolumn{6}{c|}{$\langle n_{sbj} \rangle$  for $\sqrt{s} = 141$ GeV}\\
    \cline{2-7}
    {\large $y_{cut}$} & \multicolumn{2}{c|}{R = 0.6} & \multicolumn{2}{c|}{R = 0.8} & \multicolumn{2}{c|}{R = 1}\\
     \cline{2-7}
       & PYTHIA & RAPGAP & PYTHIA & RAPGAP & PYTHIA & RAPGAP\\
 \hline 
 0.0005 & 4.41 & 4.52 & 4.30 & 4.40 & 4.17 & 4.26 \\
 \hline 
 0.001 & 3.70 & 3.80 & 3.54 & 3.63 & 3.39 & 3.47 \\
 \hline 
 0.003 & 2.68 & 2.76 & 2.50 & 2.57 & 2.36 & 2.42 \\
 \hline 
 0.005 & 2.28 & 2.35 & 2.11 & 2.17 & 1.98 & 2.04 \\
 \hline 
 0.01 & 1.81 & 1.87 & 1.67 & 1.72 & 1.57 & 1.61 \\
 \hline 
 0.03 & 1.29 & 1.31 & 1.22 & 1.23 & 1.19 & 1.20 \\
 \hline 
 0.05 & 1.14 & 1.15 & 1.11 & 1.11 & 1.10 & 1.10 \\
 \hline 
 0.1 & 1.04 & 1.04 & 1.03 & 1.03 & 1.03 & 1.03 \\
 \hline 
 
    \end{tabular}}
    \caption{Number of subjets within a jet  with a minimum  $E_{T,jet}$ cut for multiple values of $y_{cut}$ and jet radius R. $10^{7}$ events are generated at each $\sqrt{s}$.~Statistical errors are found to be negligible and excluded from the table.}
    \label{tab:1}
\end{table}

\begin{table}[h!]
    \centering
    \scalebox{0.80}{
    \scriptsize
    \begin{tabular}{|c|c|c|c|c|c|c|c|}
    \hline
       &\multicolumn{7}{c|}{$\langle n_{sbj} \rangle$  for $\sqrt{s} = 300$ GeV}\\
    \cline{2-8}
    {\large $y_{cut}$} & \multicolumn{2}{c|}{R = 0.6} & \multicolumn{2}{c|}{R = 0.8} & \multicolumn{2}{c|}{R = 1} & \\
     \cline{2-7}
       & PYTHIA & RAPGAP & PYTHIA & RAPGAP & PYTHIA & RAPGAP & Experiment \cite{zeusmain}\\
 \hline 
 0.0005 & 4.41 & 4.52 & 4.30 & 4.40 & 4.17 & 4.26 & 4.30 $\pm$ 0.0068\\
 \hline 
 0.001 & 3.70 & 3.80 & 3.54 & 3.63 & 3.39 & 3.47 & 3.49 $\pm$ 0.0057\\
 \hline 
 0.003 & 2.68 & 2.76 & 2.50 & 2.57 & 2.36 & 2.42 & 2.44 $\pm$ 0.0043\\
 \hline 
 0.005 & 2.28 & 2.35 & 2.11 & 2.17 & 1.98 & 2.04 & 2.06 $\pm$ 0.0038\\
 \hline 
 0.01 & 1.81 & 1.87 & 1.67 & 1.72 & 1.57 & 1.61 & 1.63 $\pm$ 0.0033\\
 \hline 
 0.03 & 1.29 & 1.31 & 1.22 & 1.23 & 1.19 & 1.20 & 1.20 $\pm$ 0.0024\\
 \hline 
 0.05 & 1.14 & 1.15 & 1.11 & 1.11 & 1.10 & 1.10 & 1.10 $\pm$ 0.0018\\
 \hline 
 0.1 & 1.04 & 1.04 & 1.03 & 1.03 & 1.03 & 1.03 & 1.03 $\pm$ 0.0010\\
 \hline 
    \end{tabular}}
    \caption {Number of subjets within a jet  with a minimum $E_{T,jet}$ cut for different values of $y_{cut}$ and jet radius R = 0.6 , 0.8, 1.0 for HERA data at $\sqrt{s}$= 300 GeV. $10^{7}$ events are generated for each set. Statistical errors are found to be $<$0.05\% for each value.~Results from the ZEUS experiment are shown in the last column.}
    \label{tab3}
\end{table}

\begin{figure}[ht]
\centering
\resizebox{7cm}{!}{\includegraphics{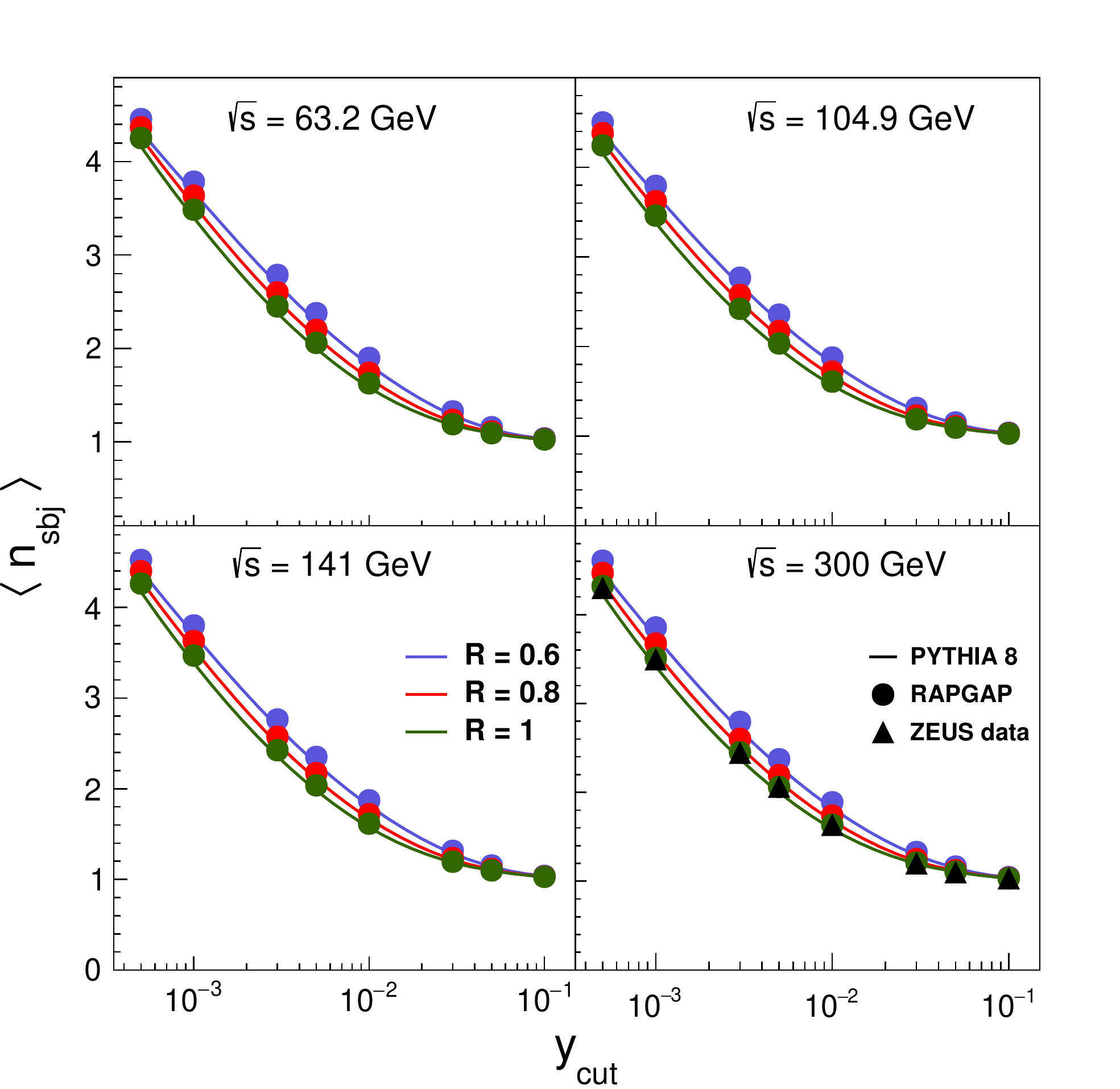}}
\caption{Mean subjet multiplicity as a function of $y_{cut}$. The errors on the simulated values are less than $0.05\%$ and the errors on the ZEUS data are very small as shown in table \ref{tab3}.}
\label{ycut_partition}
\label{ycut_partition_data}
\end{figure}

\begin{figure}[ht]
\centering
\resizebox{7cm}{!}{\includegraphics{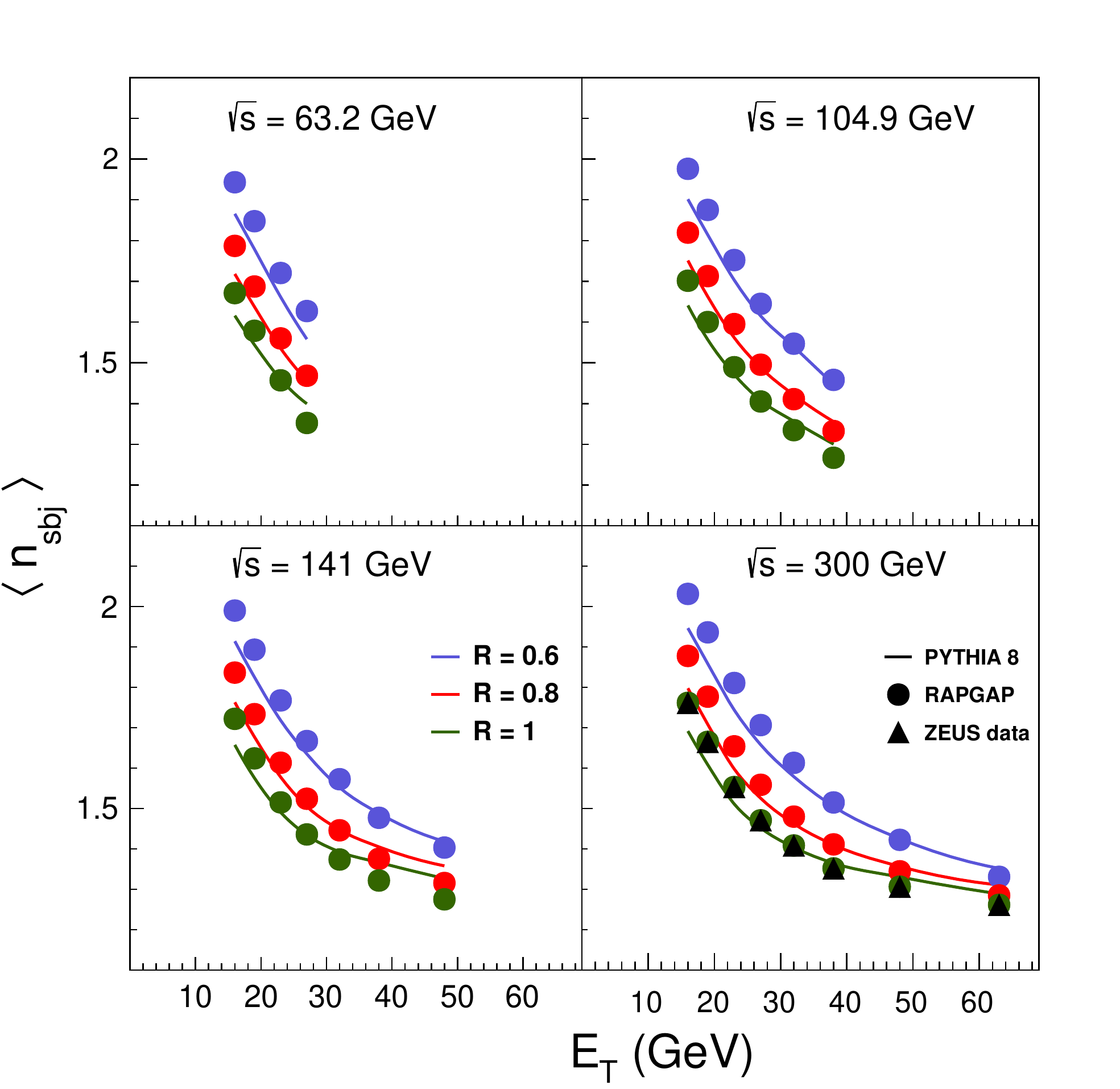}}
\caption{Mean subjet multiplicity as a function of $E_{T,jet}$.~Solid lines represent the predictions from PYTHIA and points represent predictions from RAPGAP for different values of R.} 
\label{Et_partition}
\end{figure}

For each center of mass energy, Figure \ref{ycut_partition} shows mean subjet multiplicity $\langle n_{sbj}\rangle$ as a function of $y_{cut}$ at each radius. It is seen that mean subjet multiplicity decreases rapidly with increase in value of $y_{cut}$. It can be coherently observed that for a given $y_{cut}$ value, mean subjet multiplicity $\langle n_{sbj}\rangle$ reduces slightly as the jet radius R increases. The figure also shows a comparison of the data at $\sqrt{s}$ = 300 GeV from the ZEUS experiment\cite{zeusmain} with the data simulated from PYTHIA 8 and RAPGAP. A good agreement is observed.
Figure \ref{Et_partition} presents mean subjet multiplicity as a function of jet transverse energy $E_{T,jet}$ at $y_{cut}$=10$^{-2}$. Average subjet multiplicity is observed to decrease as $E_{T,jet}$ increases for each center of mass energy. Similar observation was also made with data at $\sqrt{s}$ = 300 GeV which agrees very well with the simulated data with jet of radius = 1 in the paper \cite{zeus1999measurement, gonzalez2002measurement}. In this paper, jet shape was used to study jet substructure with conclusion that jets become narrower as $E_{T,jet}$ increases, resulting in lower values of subjet multiplicity.
\subsection{Comparison of subjets from \texorpdfstring{$k_{T}$}{TEXT} and \texorpdfstring{anti-$k_{T}$}{TEXT} algorithms}
\begin{figure}[h]
\centering
\resizebox{7cm}{!}{\includegraphics{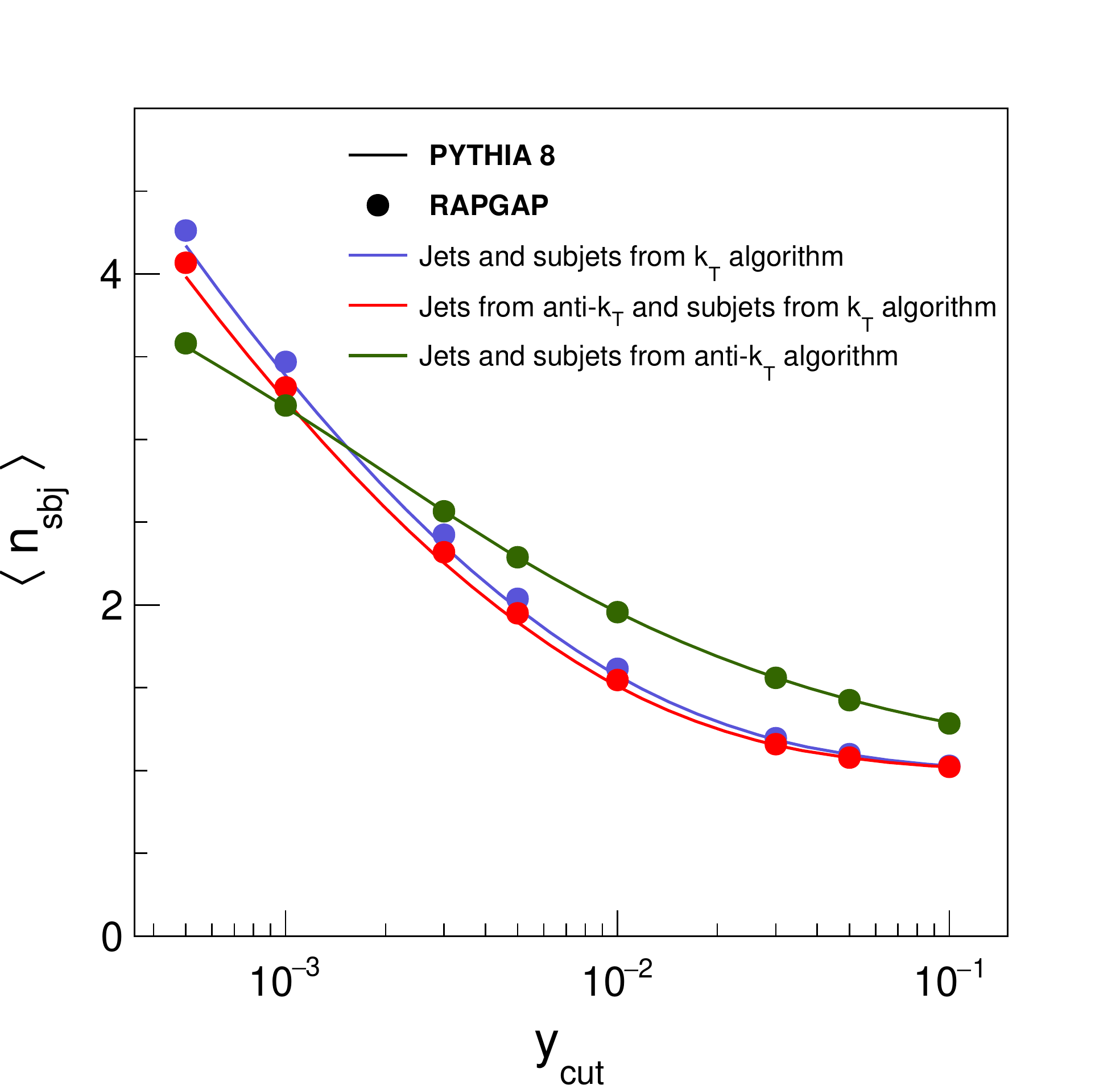}}
\caption{The mean subjet multiplicity as a function of $y_{cut}$ for $\sqrt{s}$ = 141 GeV for R = 1 and $R_{sub}$ = 0.4.}
\label{04Rs_eic3_ycut_comparison}
\end{figure}

\begin{figure}[ht]
\centering
\resizebox{7cm}{!}{\includegraphics{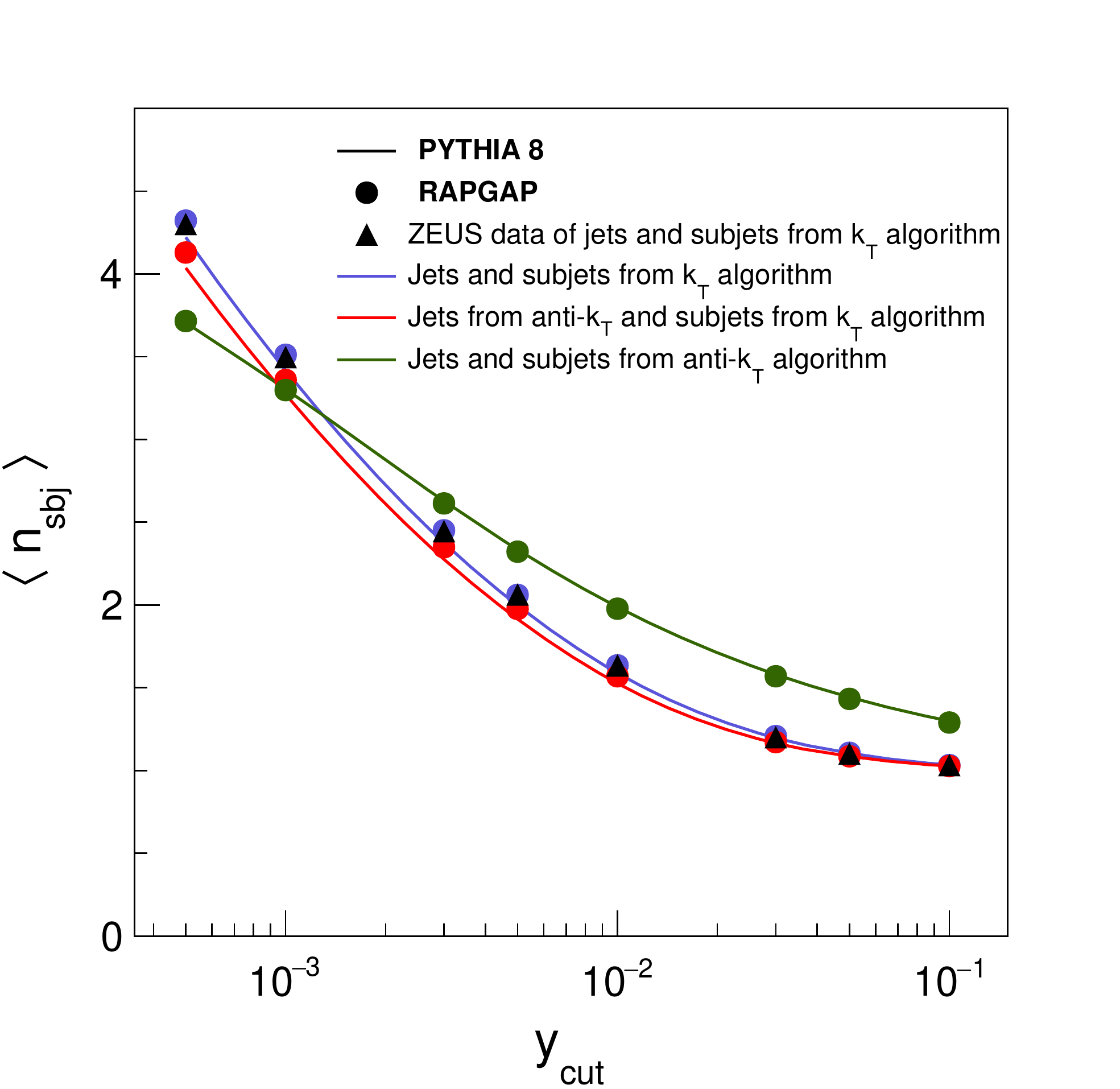}}
\caption{The mean subjet multiplicity as a function of $y_{cut}$ for $\sqrt{s}$ = 300 GeV for R = 1 and $R_{sub}$ = 0.4. The errors shown in table \ref{tab3} for ZEUS data are negligibly small.}
\label{04Rs_hera_ycut_comparison}
\end{figure}

Comparison of $k_{T}$ and anti-$k_{T}$ algorithms is studied on the production of jets and subjets for DIS $e^{+}p$ events with $E_{T,min} > $ 10 GeV and $Q^{2} >$ 125 GeV$^2$. In both the cases when jets are produced by $k_{T}$ or anti-$k_{T}$ algorithm for R =1, subjets are produced through  the $k_{T}$ algorithm for $E_{T,jet} >$ 15 GeV and -1 $< \eta_{jet} <$ 2 and the subjet multiplicity is measured. The results are shown in Figure \ref{04Rs_eic3_ycut_comparison} for $\sqrt{s}$=141 GeV and it is found that in both the cases for the chosen jet radius and $y_{cut}$, values are nearly the same for both PYTHIA and RAPGAP.~A similar study is also presented in Figure \ref{04Rs_hera_ycut_comparison} for $\sqrt{s}$=300 GeV and compared to the data from the ZEUS experiment \cite{zeusmain}.~It is observed that the data also agrees well with the predicted values from PYTHIA and RAPGAP.

In a second study, jets are produced from anti-$k_{T}$ algorithm and subjets are again produced using anti-$k_{T}$ algorithm with smaller subjet radius of  $R_{sub}$ = 0.4 \cite{bell2016mono} at both the cms energies. It is observed that the $\langle n_{sbj}\rangle$ dependence on $y_{cut}$ does not follow the trend as in the case when subjets are formed from $k_{T}$ algorithm, as evident from Figures \ref{04Rs_eic3_ycut_comparison} and \ref{04Rs_hera_ycut_comparison}.~It is observed that the $\langle n_{sbj}\rangle$ decreases less steeply with $y_{cut}$ in this case.~A further investigation in to this observation reveals that equation \ref{jet_algorithm} is dominated by high $p_{T}$ for anti-$k_{T}$ algorithm ($\rho = -1$) and tends to cluster hard radiation first.~This assigns majority of energy to one subjet and the weaker subjet is most likely to get discarded.~This leads to the production of imbalanced subjets.~But $k_{T}$ algorithm prefers to cluster soft particles first, which results in even distribution of energy between subjets that contain final state radiation \cite{krohn2010jet}.~The deviation observed is also reported in \citep{atkin2015review} which states that anti-$k_{T}$ algorithm prefers to cluster hard particles first, which makes it best at resolving jets but is ineffective for producing jet substructure due to its inability to de-cluster.
\subsection{Differential jet shape measurements}
\begin{figure}[ht]
\centering
\resizebox{7cm}{!}{\includegraphics{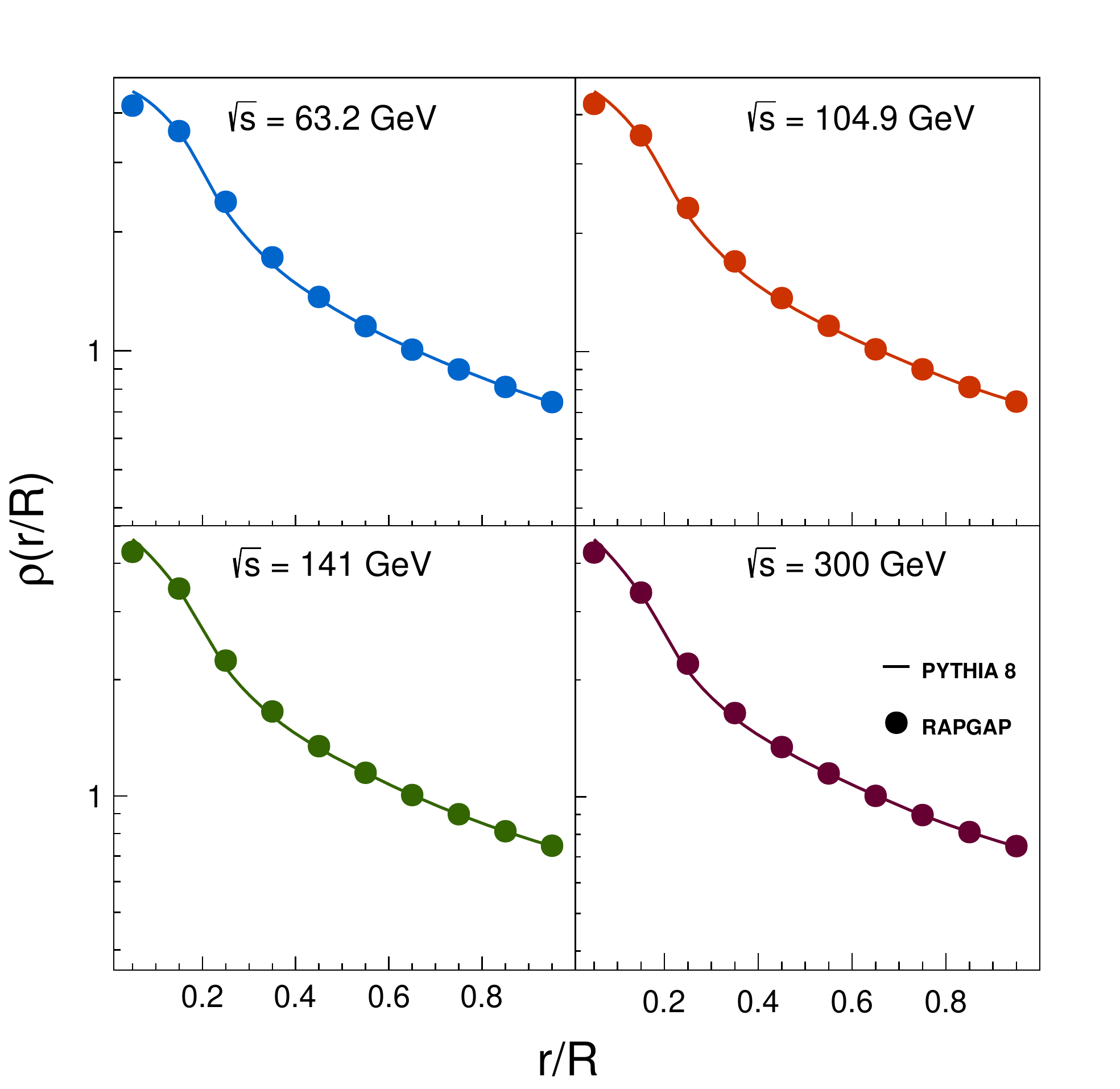}}
\caption{Differential jet shape as the function of annulus radii $r$.}
\label{jetshape_partition}
\end{figure}

For differential jet shape measurements, jets are produced through longitudinally invariant $k_T$ algorithm for jet radius R = 1. Figure \ref{jetshape_partition} shows measurement of differential jet shape as defined in equation \ref{equation_jetshape} at various annulus radii $r$ for each energy.~An apparent peak is observed in differential jet shape nearby the jet-axis, indicating the presence of high $E_{T}$ region around center of the jet.~The differential jet shape variable $\rho$ decreases as $r$ increases.~For a specific annulus radius, jet shape is observed to be similar at each center of mass energy for both PYTHIA and RAPGAP.~This leads to the observation that both the subjet multiplicity and jet shape are independent of center of mass energy.

\subsection{Experimental outlook}
\begin{figure}[h]
\centering
{\resizebox{2.8cm}{!}{\includegraphics{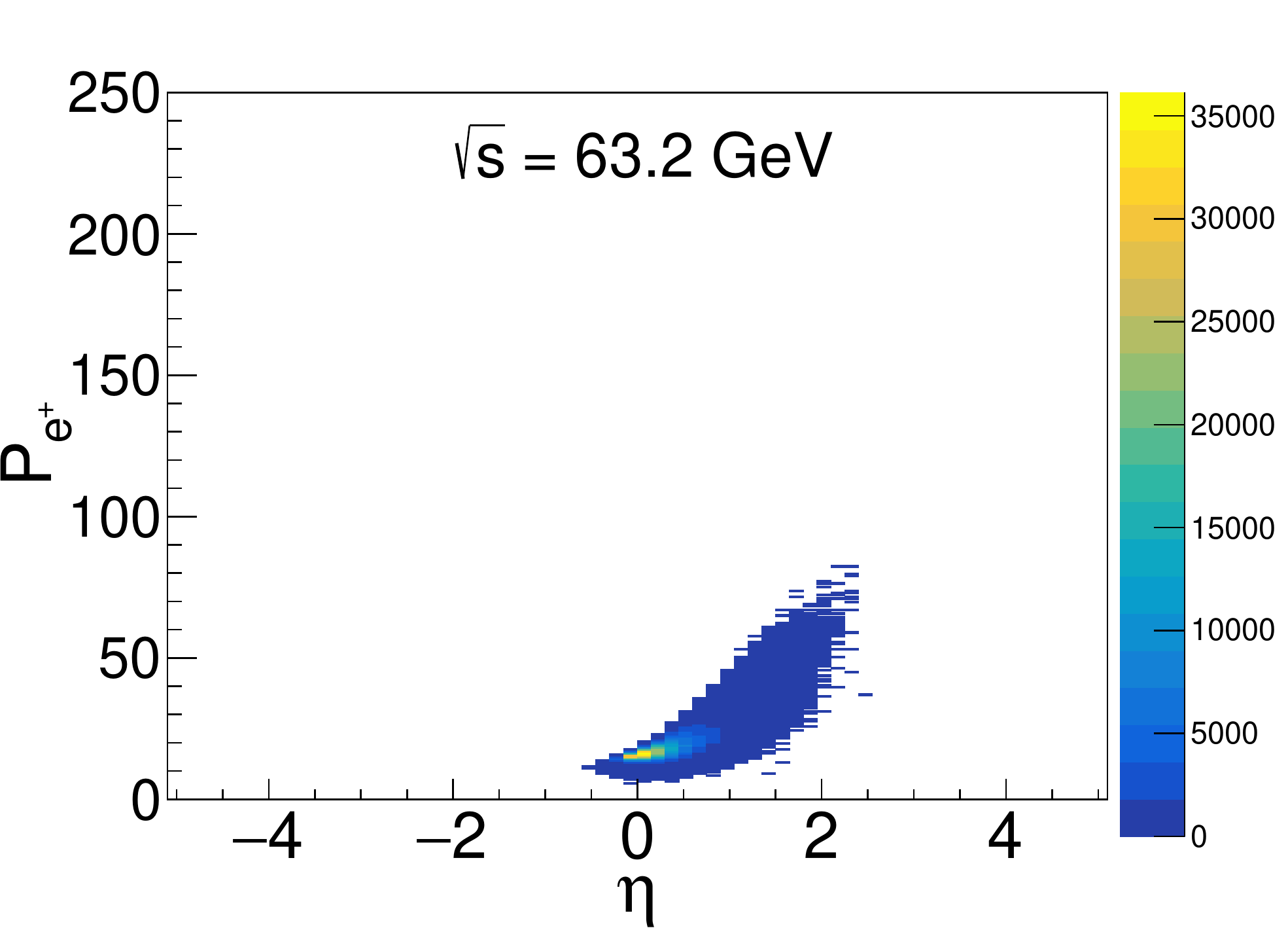}}
\resizebox{2.8cm}{!}{\includegraphics{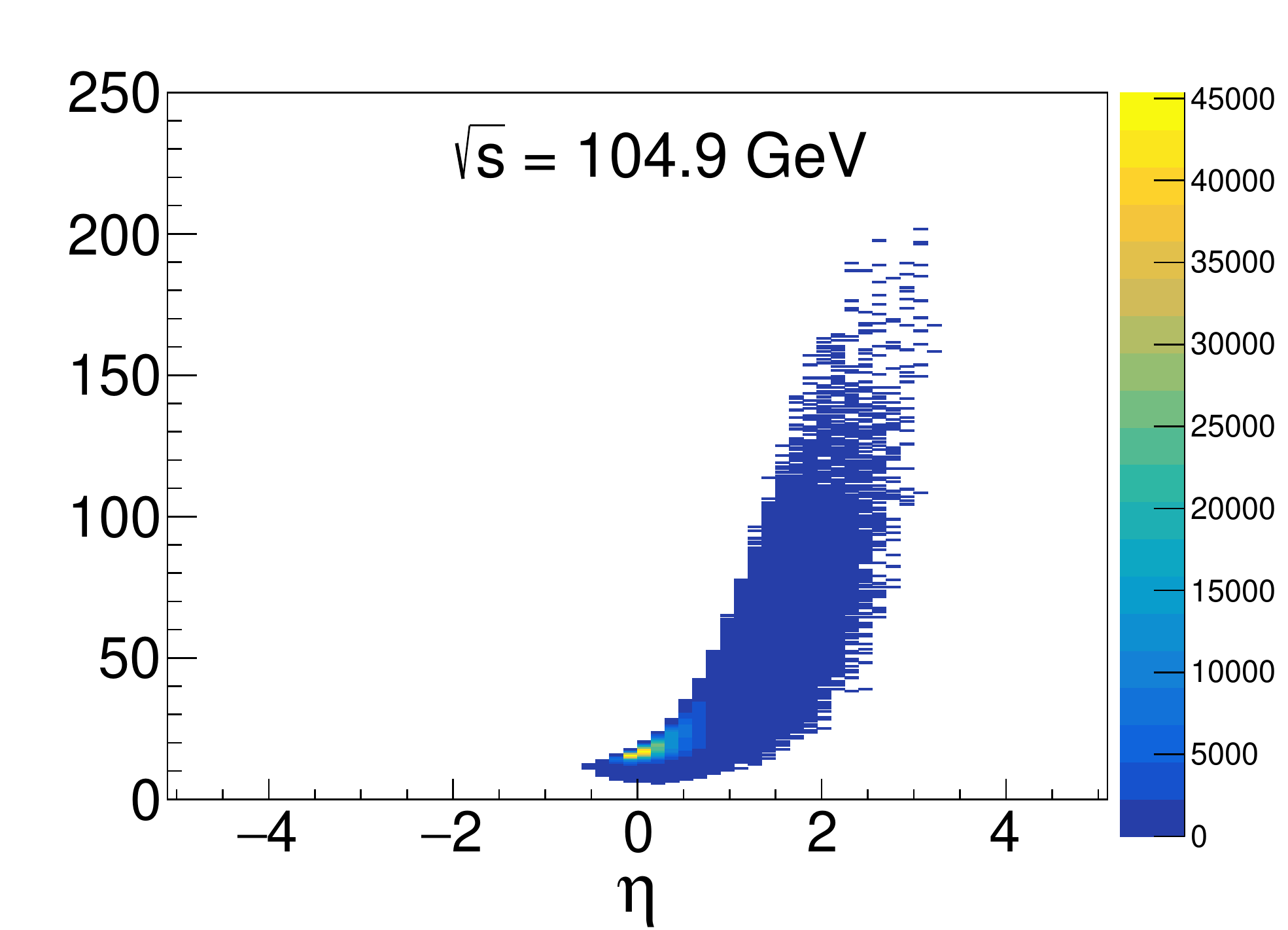}}
\resizebox{2.8cm}{!}{\includegraphics{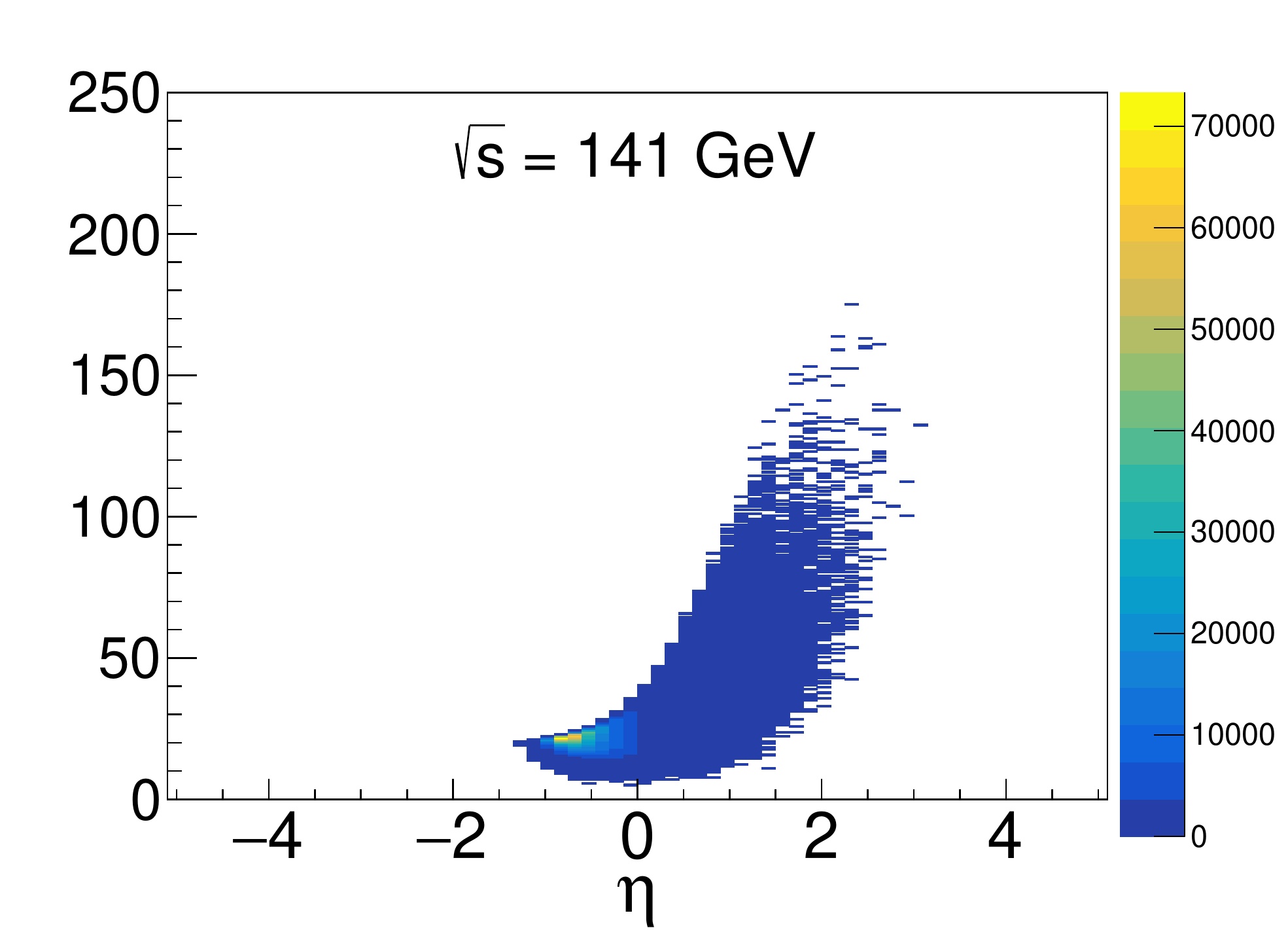}}
\resizebox{2.8cm}{!}{\includegraphics{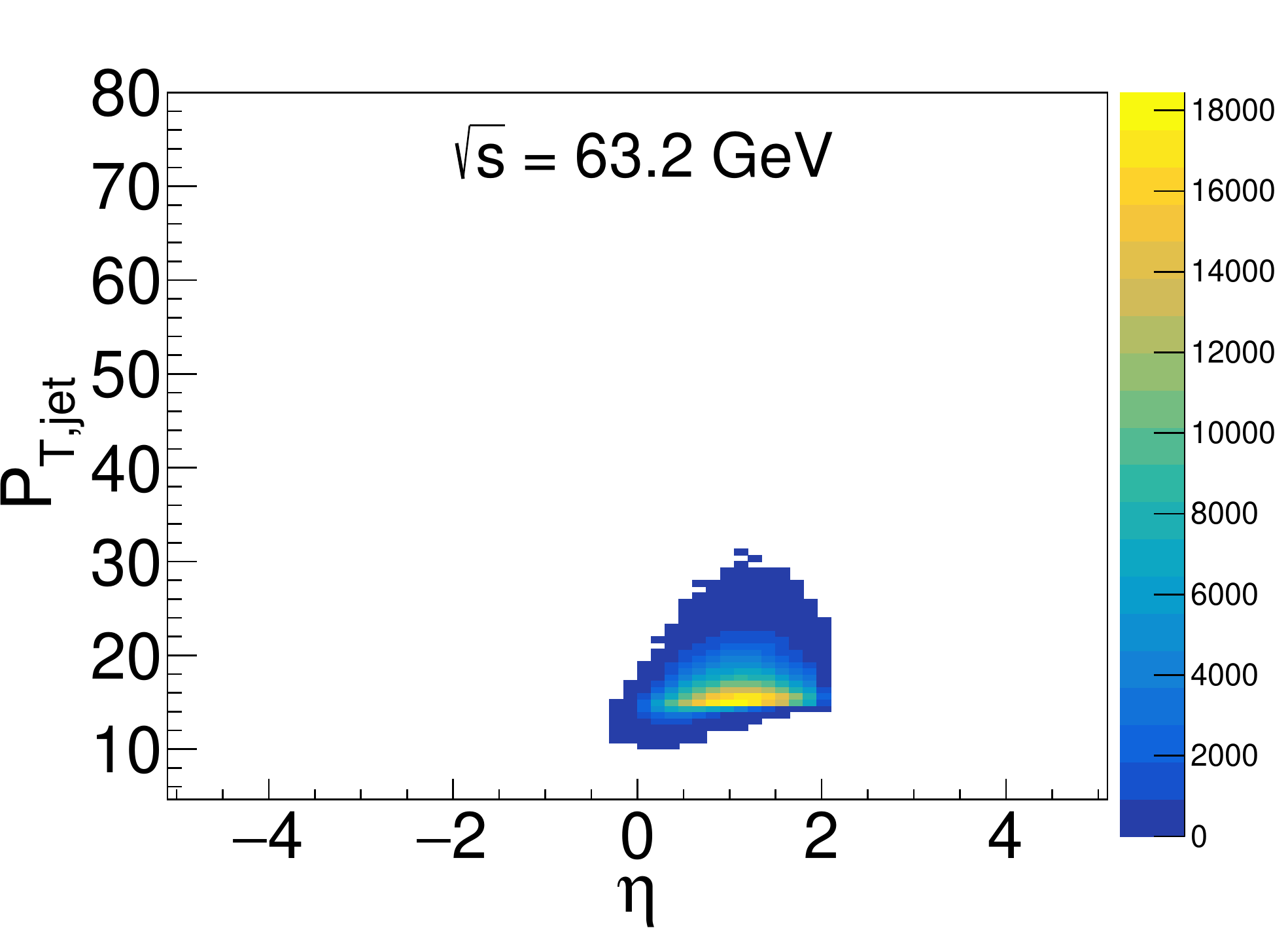}}
\resizebox{2.8cm}{!}{\includegraphics{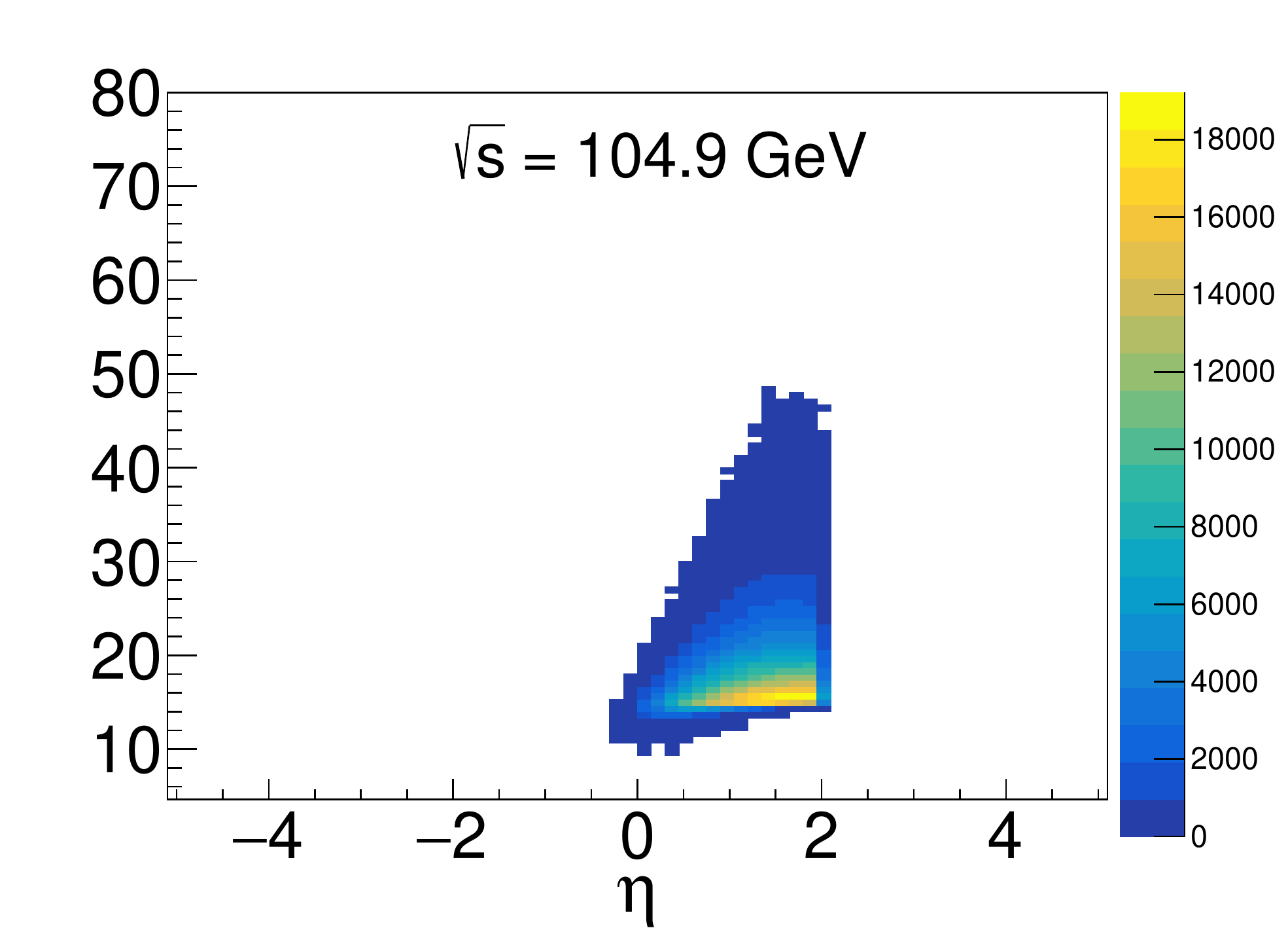}}
\resizebox{2.8cm}{!}{\includegraphics{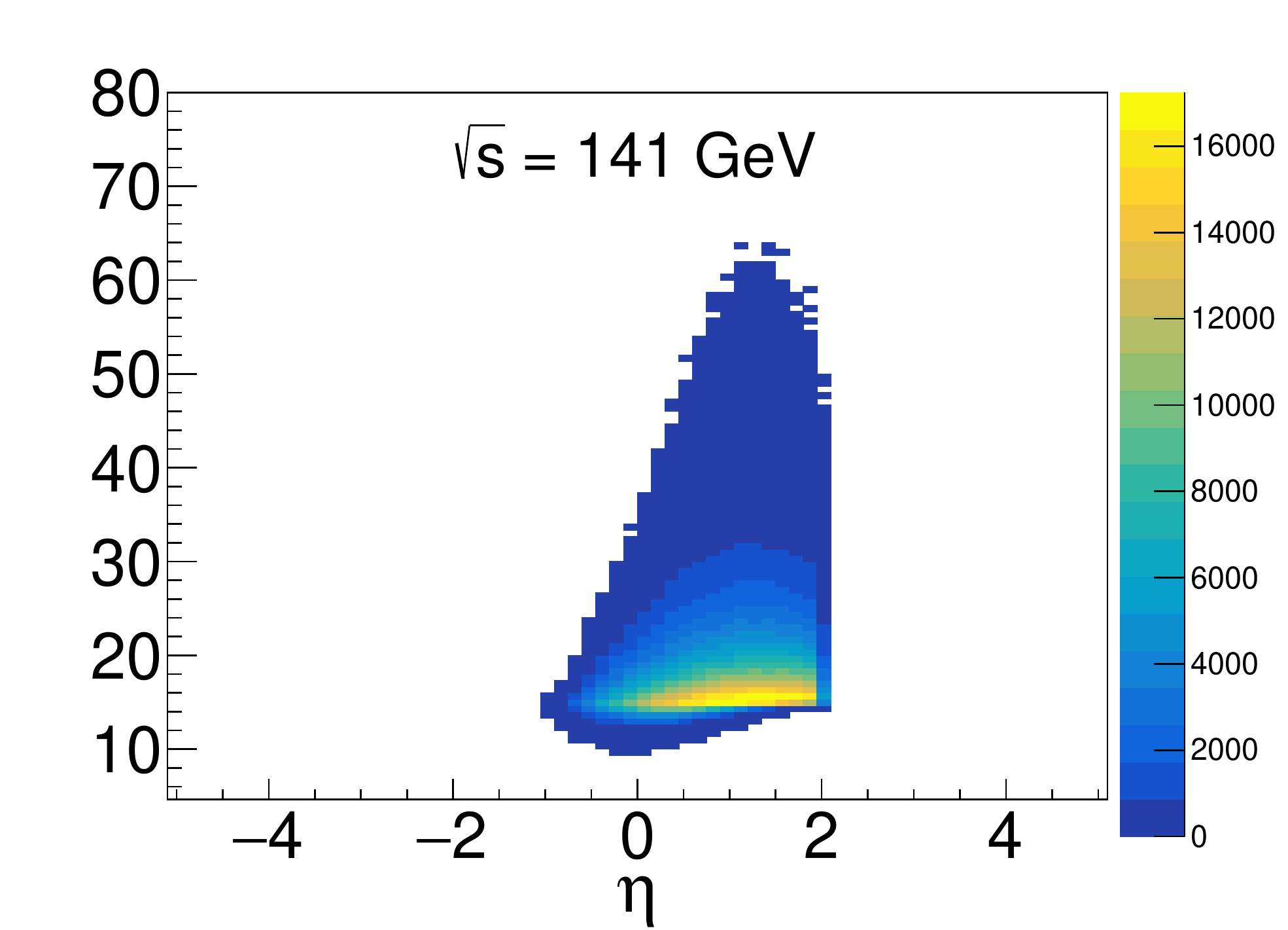}}}
\caption{Positron momentum ($P_{e^+}$) and jet transverse momentum ($P_{T,jet}$) as function of their respective pseudorapidity $\eta$, in the events with $E_{T,jet} > $ 15 GeV and $-1<\eta_{jet}<$ 2.}
\label{positron_fig}
\end{figure}

The jet and jet-substructure studies at EIC in $ep$ collisions would serve as one of the most important calibration tools for performing $eA$ studies.~The jets in the $eA$ collisions traverse through the nuclear medium. Differences between the jet-substructure in $ep$ and $eA$ collisions  provide information of how partonic energy is dissipated in the cold nuclear medium.~High $P_T$ jets are expected to be present mostly in the region $-1 < \eta < 2$, where the scattered positron is also present as seen in Figure \ref{positron_fig} which shows the positron momentum as a function of its pseudorapidity. It can be seen that for $\sqrt{s}$ = 63.2, 104.9, and 141 GeV the highest positron momentum reaches around 90, 200, and 180 GeV/c and jet transverse momentum is found up to 30, 50, 65 GeV/c respectively. The positron position in the detector is $Q^2$ dependent and for the $Q^2 > 125$~GeV$^2$, it is found in the barrel and forward endcap region. 
 Higher center-of-mass energies would be vital to enable the jet studies at EIC.~The precision of jet energy and positron energy measurements should be better than the earlier experiments at HERA. As the luminosity increases manifold at EIC, there is a need to improve the detector related systematic errors. The increased luminosity and hence the increased event rate are an asset at the EIC which dictates a high data rate of $\approx$500 kHz \cite{willeke2021electron}.~For the informative jet substructure studies, the lepton-hadron identification is crucial in the barrel and forward endcap regions.~The positron measurements are designed using cylindrical symmetric electromagnetic detectors in the barrel region and block shaped electromagnetic detectors in the forward region.

The hadron physics can be handled using  high granularity electromagnetic and hadronic calorimeters.~A detailed discussion on the  optional technologies and the region covered by them can be found in \cite{willeke2021electron}.
\section{CONCLUSION}
Predictions for the properties of subjets within jets produced in neutral current deep inelastic $e^{+}p$ scattering at the future EIC are presented using simulated data.~The mean subjet multiplicity at each of the center-of-mass energies of 63.2, 104.9 and 141 GeV has been found using jets in the pseudorapidity region, $-1<\eta_{jet}<$ 2 and with each jet having $E_{T,jet} > $ 15 GeV.~Various combinations of jet clustering algorithms are used and compared.~It is observed that the average number of subjets within a jet decreases as $E_{T,jet}$ increases.~The observation agrees with the results from HERA.

It has been shown that for jets, both $k_{T}$ and anti-$k_{T}$ algorithms give similar results.~Multiple values of the jet radius are used for selecting jets and subjets.~It is observed that the subjet multiplicity decreases as the jet radius increases.~For each radius, subjet multiplicity is measured for various values of $y_{cut}$, the resolution parameter.~Subjet multiplicity is observed to decrease with the increase in $y_{cut}$.~The jet shape is studied in terms of differential jet shape variable $\rho(r)$, which is observed to decrease as the annulus radius is increased.~It has been shown that both subjet multiplicity and differential jet shape in a jet of given transverse energy  $E_{T}$, are independent of the center of mass energy.~When jets and subjets both are produced by $k_{T}$ algorithm, mean subjet multiplicity tends to decrease as $E_{T,jet}$ is increased.

Comparison of $k_{T}$ algorithm and anti-$k_{T}$ algorithm has been studied for the production of subjets. It is observed that for the subjets, $k_{T}$ algorithm is much better suited than anti-$k_{T}$ algorithm as can be well observed from the Figure \ref{04Rs_hera_ycut_comparison} for the data at $\sqrt{s}$ = 300 GeV.~It is due to the fact that anti-$k_{T}$ algorithm prefers to cluster hard particles first, which makes it best at resolving jets but due to its inability to decluster, it is ineffective for producing substructure.~It is observed that characteristics of jets and subjets produced in PYTHIA and RAPGAP are in close agreement.

The upcoming EIC detector can increase the accessible range in $x$ for a given $Q^2$ due to its reduced cms energy as compared to the HERA detector.~This region is interesting and not much explored.~The improved  precision of the proton parton distributions is foreseen at the EIC detector due to the promising increase in the luminosity.~The jet and jets' substructure studies at the EIC would have very small statistical uncertainties and a deep knowledge of the detector would be required to pin down the systematic uncertainties.

\bibliography{subjet.bib}
\end{document}